\documentclass[a4paper,12pt]{article}
\pdfoutput=1
\usepackage[DIV=14,BCOR=0mm,headinclude=true,footinclude=false]{typearea}
\usepackage[utf8]{inputenc}
\usepackage[T1]{fontenc}

\usepackage{style}

\usepackage{subcaption}
\usepackage{tipa}
\usepackage{graphicx}
\usepackage[bottom]{footmisc}
\usepackage[titletoc]{appendix}
\usepackage{hyperref}
\usepackage{listings}
\usepackage{float}
\usepackage{amsmath}
\usepackage{mathrsfs}
\usepackage{physics}
\usepackage{amssymb}
\usepackage{mathtools}
\usepackage{amsfonts}
\usepackage{braket}
\usepackage{slashed}
\usepackage{dashbox}
\usepackage[colorinlistoftodos]{todonotes}
\usepackage{tikz-feynman}   
\usepackage{tikz}
\usetikzlibrary{snakes}     
\usepackage[normalem]{ulem}

\usepackage{enumitem}       

\usepackage{scalerel}

\usepackage{empheq}         
\usepackage[many]{tcolorbox} 
\tcbset{
	highlight math style={
		colframe=black,
		colback=white,
		arc=2pt,
		boxrule=0.7pt,
	}
}

\usepackage{pict2e}         

\makeatletter
\DeclareRobustCommand{\loplus}{\mathbin{\mathpalette\dog@lsemi{+}}}
\DeclareRobustCommand{\lotimes}{\mathbin{\mathpalette\dog@lsemi{\times}}}
\DeclareRobustCommand{\roplus}{\mathbin{\mathpalette\dog@rsemi{+}}}
\DeclareRobustCommand{\rotimes}{\mathbin{\mathpalette\dog@rsemi{\times}}}

\newcommand{\dog@rsemi}[2]{\dog@semi{#1}{#2}{-90,90}}
\newcommand{\dog@lsemi}[2]{\dog@semi{#1}{#2}{270,90}}
\newcommand{\dog@semi}[3]{%
	\begingroup
	\sbox\z@{$\m@th#1#2$}%
	\setlength{\unitlength}{\dimexpr\ht\z@+\dp\z@\relax}%
	\makebox[\wd\z@]{\raisebox{-\dp\z@}{%
			\begin{picture}(1,1)
				\linethickness{\variable@rule{#1}}
				\roundcap
				\put(0.5,0.5){\makebox(0,0){\raisebox{\dp\z@}{$\m@th#1#2$}}}
				\put(0.5,0.5){\arc[#3]{0.5}}
			\end{picture}%
	}}%
	\endgroup
}
\newcommand{\variable@rule}[1]{%
	\fontdimen8  
	\ifx#1\displaystyle\textfont3\else
	\ifx#1\textstyle\textfont3\else
	\ifx#1\scriptstyle\scriptfont3\else
	\scriptscriptfont3\relax
	\fi\fi\fi
}
\makeatother

\newcommand\reallywidehat[1]{\arraycolsep=0pt\relax%
	\begin{array}{c}
		\stretchto{
			\scaleto{
				\scalerel*[\widthof{\ensuremath{#1}}]{\kern-.5pt\bigwedge\kern-.5pt}
				{\rule[-\textheight/2]{1ex}{\textheight}} 
			}{\textheight} %
		}{0.5ex}\\           
		#1\\                 
		\rule{-1ex}{0ex}
	\end{array}
}

\def\be{\begin{equation}}
	\def\ee{\end{equation}}
\def\ba{\begin{aligned}}
	\def\ea{\end{aligned}}

\newcommand{\SL}{\text{SL}(2,\mathbb{R})}

\newcommand{\hor}{\mathscr{H}}

\numberwithin{equation}{section}
\numberwithin{table}{section}

\title{Taming the Aretakis instability: extremal black holes with multi-degenerate horizons}

\author[a,b]{Shreyansh Agrawal,\footnote{\texttt{sagrawal@sissa.it}}}
\author[a,b]{Panagiotis Charalambous,\footnote{\texttt{pcharala@sissa.it}}}
\author[a,b]{Laura Donnay,\footnote{\texttt{ldonnay@sissa.it}}}
\author[a,b,c]{Stefano Liberati,\footnote{\texttt{liberati@sissa.it}}}
\author[a,b,c]{and Giulio Neri\footnote{\texttt{gneri@sissa.it}}}

\affiliation[a]{International School for Advanced Studies (SISSA), \\
Via Bonomea 265, 34136 Trieste, Italy}
\affiliation[b]{National Institute for Nuclear Physics (INFN), \\
Sezione di Trieste, Via Valerio 2, 34127, Italy}
\affiliation[c]{Institute for Fundamental Physics of the Universe (IFPU), \\
Via Beirut 2, 34014 Trieste, Italy}

\abstract{
    Stationary black hole geometries with non-degenerate Cauchy horizons are classically unstable due to mass inflation. At extremality, mass inflation is absent, but a different dynamical instability arises: the Aretakis instability. In this work, we investigate the properties of degenerate horizons and their associated Aretakis instabilities. By studying examples with increasingly higher-order horizon degeneracy, we show that the Aretakis instability weakens as the degree of degeneracy grows. Motivated by these results, we propose a new black hole geometry characterized by an infinitely degenerate horizon, which we argue is stable under Aretakis-type perturbations and may therefore provide a concrete realization of a ``graveyard'' end state for these objects.
}

\begin{document}

\maketitle


\section{Introduction}
\label{sec:Intro}

Over the years, considerable attention has been devoted to black hole exteriors and event horizons. For non-extremal solutions, the latter enjoy well-established classical stability properties, which are closely tied to the redshift effect (see, e.g., \cite{Dafermos:2005eh,Dafermos:2008en,Aretakis:2010gd} and references therein). At the semiclassical level, Hawking radiation leads to a gradual decrease of the horizon area, an ultra-slow (yet theoretically significant) process for astrophysical black holes.

By contrast, black hole interiors have recently come under renewed scrutiny, largely driven by growing interest in quantum gravity, regular (singularity-free) solutions, and their possible phenomenological viability (see, e.g., \cite{Carballo-Rubio:2025fnc} for a comprehensive overview). Understanding the stability properties of black hole interiors, whether singular or non-singular, remains a central challenge at both the classical and semiclassical levels.

Inner horizons, in particular, lie at the crossroads of several dynamical phenomena and host a variety of instabilities whose origin and severity depend sensitively on the underlying geometric structure. More broadly, a number of such instabilities have emerged as especially relevant to the question of whether a black hole can maintain quasi-stationary behavior under perturbations. Since inner horizons are present not only in regular black hole geometries but also in astrophysically relevant Kerr black holes, clarifying their role as a potential source of instability has become increasingly urgent.

Spacetimes with non-extremal inner horizons are well known to suffer from the so-called mass inflation instability~\cite{Simpson:1973ua,Poisson:1989zz,Poisson:1990eh,Bonanno:1994ma,Bonanno:1994qh,Brown:2011tv,Frolov:2017rjz,Carballo-Rubio:2018pmi,Mcmaken:2021isj,Visser:2024zkx}, a mechanism that has recently been shown to affect slowly evolving inner horizons as well~\cite{Carballo-Rubio:2024dca}. In addition, a distinct semiclassical instability arises at the level of the renormalized stress-energy tensor~\cite{Hollands:2019whz}. In stationary geometries, this can be understood as a consequence of the fact that the standard Unruh state cannot be regular at both the outer and inner horizons whenever the squared surface gravities of the event and Cauchy horizons differ, $\kappa_{\mathrm{in}}^{2} \neq \kappa_{\mathrm{out}}^{2}$~\cite{McMaken:2023uue,McMaken:2024fvq}.

Although the ultimate endpoint of these instabilities remains uncertain, there is now compelling evidence, from both analytical studies~\cite{Barcelo:2020mjw,Barcelo:2022gii} and numerical investigations~\cite{Barenboim:2025ckx,Boyanov:2025otp}, that the semiclassical instability at the inner horizon may dominate over mass inflation and possibly drive a complete evaporation of the trapped region from within. If so, this would provide a strong dynamical push toward an extremal configuration.

Indeed, it is natural to suspect that an extremal geometry might represent an endpoint of the instability, since in that case the spacetime possesses a single degenerate horizon with vanishing surface gravity. For a stationary geometry in which horizons are determined by the roots of a metric function $f(r)=0$, a degenerate horizon is characterized by the additional condition $f'(r)=0$ at the same location. In the presence of a singularity, one may further argue that the evolution cannot proceed beyond extremality without destroying the horizon altogether and thereby violating weak cosmic censorship, i.e.\ by producing a naked singularity. For regular black holes (see, e.g., \cite{Bardeen1968,Dymnikova1992,Hayward2006}), this obstruction is absent, since the horizonless configuration beyond extremality would simply describe an ultra-compact object with light rings. Nevertheless, this possibility is itself problematic, because such light rings are expected to occur in pairs~\cite{Cunha:2017qtt,DiFilippo:2024ddg}: an outer unstable one, as in the black hole case, and an inner stable one.

The issue is that linear perturbations around stable light rings decay extremely slowly~\cite{Keir:2014oka}, potentially leading to a nonlinear instability~\cite{Cardoso:2014sna,Cunha:2017qtt}. Although it remains unclear how efficient and generic this mechanism is, it adds to other open problems affecting black hole mimickers such as energy absorption at the surface and the stability of ergoregions without horizons in rapidly rotating configurations, and therefore strengthens the case for considering extremal black holes as possible stable endpoints of these processes.

It is in this context that the relevance of the Aretakis instability, uncovered for extremal horizons \cite{Aretakis:2011ha,Aretakis:2011hc}, becomes clear, not merely as a feature of mathematical relativity, but as a potential obstruction to the stability of degenerate horizons. In brief, the Aretakis instability causes certain transverse derivatives of massless scalar fields and, more generally, of higher-spin fields \cite{Lucietti:2012sf,Lucietti:2012xr,Apetroaie:2022rew}, to grow without bound along the horizon. This effect has been studied extensively for the most relevant extremal black holes in General Relativity, namely the extremal Reissner--Nordstr\"om and Kerr solutions. However, it appears to be considerably more generic, since it depends only on the near-horizon structure of the metric \cite{Gralla:2016jfc,Gralla:2017lto,Gralla:2018xzo}.

To assess whether the Aretakis instability is an unavoidable feature of degenerate horizons, in this work we extend its analysis to black holes with multiply degenerate horizons, namely horizons at which not only the metric function $f(r)$ and its first derivative vanish, but also higher derivatives, up to some finite order. Such configurations were first proposed in the context of non-extremal regular black hole geometries possessing a degenerate inner horizon, since the vanishing of the corresponding surface gravity removes the mass inflation instability \cite{Carballo-Rubio:2022kad,Franzin:2022wai}. The same idea was later generalized to construct black hole metrics with multiple, non-coincident, multiply degenerate horizons. Because the surface gravity vanishes at each horizon, these geometries may be stable against mass inflation, semiclassical instabilities at the inner horizon, and Hawking evaporation. For this reason, they were dubbed ``black hole graveyards'' \cite{DiFilippo:2024spj}.

As already emphasized in \cite{DiFilippo:2024spj}, however, no definitive conclusion about the stability of such multi-degenerate black hole solutions can be reached without first assessing their behavior with respect to the Aretakis instability. Do multiply degenerate horizons suffer from the same instability as the standard degenerate horizons associated with extremal black holes? This is the central question addressed in the present paper. As we shall show, multiply degenerate horizons are in fact more stable than simply degenerate ones, with the instability becoming progressively weaker as the degree of degeneracy increases. We also propose a geometry with an ``infinitely degenerate'' horizon: a geometry for which $f(r)$ and all higher derivatives vanish at the horizon, which may hence represent the ultimate ``graveyard'' solution.

The paper is organized as follows. In Section \ref{sec:AretakisReview}, we begin by revisiting the canonical example of the Aretakis instability on extremal Reissner--Nordstr\"om (ERN) black holes. A central lesson of this system is the universality of the effect: the instability is governed entirely by local near-horizon data and is therefore insensitive to the global features of the spacetime. This locality makes Aretakis-type behavior a powerful diagnostic for assessing the viability of extremal or near-extremal geometries.
In Section \ref{sec:MultiDegenerateBHs}, we investigate how the strength of the instability depends on the degree of horizon degeneracy. By considering horizons of higher-order (or ``higher-rank'') degeneracy, we explore how the blow-up rates and conserved quantities associated with perturbations evolve, and we examine possible correlations with the structure of geodesics in the near-horizon region.
We then propose a new class of infinitely degenerate black hole geometries that do not exhibit an Aretakis instability, thereby evading all presently known classical instabilities associated with black hole horizons. These solutions provide a novel counterexample to the expectation that extremality generically entails horizon instabilities, and may offer insight into how the classical and semiclassical stability issues discussed above can be reconciled in gravitational systems.
Section \ref{sec:NumericalResults} contains a numerical analysis, which confirms our analytic results, and in Section~\ref{sec:Discussion} we draw our conclusions.

Throughout this paper, we adopt a mostly-plus Lorentzian signature and geometrized units with $G = c = 1$. Angular coordinates on the transverse sphere are denoted by $x^{A}$ with $A = 1,\dots, d-2$, so that in four dimensions $x^{A} = (\theta,\phi)$.


\section{The Aretakis instability: A review}
\label{sec:AretakisReview}


\subsection{Extremal Reissner-Nordstr\"om black hole}

The extremal Reissner-Nordstr\"om (ERN) geometry describes an asymptotically flat, non-rotating and electrically charged black hole solution of Einstein-Maxwell field equations, whose electric charge $Q$ saturates the mass bound, $Q^2=M^2$. It is described, in advanced Eddington-Finkelstein coordinates, by the metric
\be
\label{eq:ERN_metric}
	ds_{\text{ERN}}^2 = -f(r)\dd\upsilon^2 + 2\dd\upsilon \dd r + r^2\dd\Omega_2^2\,,
\ee
where $\dd\Omega_2^2$ is the line element on $\mathbb{S}^2$ and
\be
	f(r) = \left(1-\frac{M}{r}\right)^2 \,,
\ee
with the double root at $r=M$ determining the location of the (degenerate) future event horizon $\mathscr H^+$.

In \cite{Aretakis:2011ha,Aretakis:2011hc} Aretakis studied the stability problem of the ERN for linear scalar\footnote{See \cite{Zimmerman:2016qtn} for the case of a charged scalar and \cite{Lucietti:2012sf,Lucietti:2012xr,Apetroaie:2022rew,Agrawal:2025fsv} for the study of electromagnetic and linearized gravitational perturbations.} perturbations by considering the evolution of a massless scalar field $\Phi$ in the ERN background geometry, with regular initial data specified on a spacelike hypersurface intersecting $\mathscr H^+$ and extending to (spacelike or null) infinity. He showed that, for generic initial data, transverse derivatives $\partial_r^k \Phi$ blow up at late time on $\mathscr H^+$, thereby indicating the presence of an instability.
Importantly, his results rely on the identification of an infinite set of conservation laws on $\mathscr H^+$, which we now briefly review (following mostly the notations and presentation in \cite{Agrawal:2025fsv}).

\subsubsection{Aretakis conserved quantities}
The scalar wave equation for the minimally coupled, real and massless scalar field perturbation $\Phi$ of the ERN geometry \eqref{eq:ERN_metric} is given by (after multiplying by $r^2$)
\be\label{eq:boxERN}
	\begin{gathered}
		0 = r^2\Box_{\text{ERN}}\Phi = \left[\partial_{r}\left(r-M\right)^2\partial_{r}  +2\partial_{\upsilon}r\partial_{r}r +\Delta_{\mathbb{S}^2}\right]\Phi \,,
	\end{gathered}
\ee
where $\Delta_{\mathbb{S}^2}$ is the Laplace operator on the two-sphere. Spherical symmetry of the background configuration ensures that the scalar wave operator distributes separately onto spherical harmonic modes $\Phi_{\ell m} = \int_{\mathbb{S}^2}d\Omega_2\,\bar{Y}_{\ell m}\Phi$ of orbital number $\ell$ and azimuthal number $m$. The resulting evolution equation of motion for each mode is
\be
	\partial_{\upsilon}\left(r\partial_{r}\left(r\Phi_{\ell m}\right)\right) = \frac{1}{2}\left[\ell\left(\ell+1\right) -\partial_{r}\left(r-M\right)^2\partial_{r}\right]\Phi_{\ell m}\,.
\ee
To reveal the emergence of a tower of near-horizon conservation laws, we first act with $k$ transverse derivatives, $\partial_{r}^{k}$, on this field equation and then take the near-horizon limit, $r\rightarrow M$, to obtain
\be\label{eq:DkERNWaveEq}
	\partial_{\upsilon}\left(\partial_{r}^{k}\left[r\partial_{r}\left(r\Phi_{\ell m}\right)\right]\right)\big|_{r=M} = \frac{1}{2}\left(\ell-k\right)\left(\ell+k+1\right)\left(\partial_{r}^{k}\Phi_{\ell m}\right)\big|_{r=M} \,.
\ee
Written in this form, it is straightforward to identify that a hierarchy of conserved quantities arises at each orbital number by setting $k = \ell$,
\be\label{eq:AretakisNHlimL}
	\begin{gathered}
		A_{\ell m} \coloneqq \frac{M^{\ell-1}}{\left(\ell+1\right)!}\lim_{r\rightarrow M}\partial_{r}^{\ell}\left[r\partial_{r}\left(r\Phi_{\ell m}\right)\right] \,, \\
		\quad\Rightarrow\quad \partial_{\upsilon}\,A_{\ell m} = 0 \,,\quad \ell\ge 0 \,.
	\end{gathered}
\ee
We will refer to these near-horizon constants $A_{\ell m}$ as the ``Aretakis conserved quantities''. In the literature, it is sometimes customary to use the following dimensionful Aretakis constants
\be
	H_{\ell m} =M^{-2}\lim_{r\rightarrow M}\partial_{r}^{\ell}\left[r\partial_{r}\left(r\Phi_{\ell m}\right)\right] = \frac{\left(\ell+1\right)!}{M^{\ell+1}}A_{\ell m} \,,
\ee
see, e.g.,~Theorem 1 of \cite{Aretakis:2011hc} or \cite{Lucietti:2012xr}.

\subsubsection{Aretakis instability}
The existence of the above conserved quantities is instrumental in demonstrating the emergence of a linear instability. Indeed, the dynamical instability properties of ERN for scalar perturbations, namely non-decay and blow-up for derivatives of $\Phi$ at late times, rely on the hierarchy of conservation laws of Eq.~\eqref{eq:AretakisNHlimL}, together with decay results for derivatives of $\Phi_{\ell m}$ transversal to $\hor^+$.

Aretakis derived pointwise estimates for the late time behavior of transverse derivatives of $\Phi$, showing that $\partial_{r}^{k}\Phi_{\ell m}$ decays along $\hor^+$ for $k\leq \ell$ (see Theorem 7.2 in \cite{Aretakis:2011hc}),
\be\label{eq:ERNkDrDecay}
	\left(\partial_{r}^{k\le\ell}\Phi_{\ell m}\right)\big|_{r=M} \xrightarrow{\upsilon\rightarrow\infty} 0 \,.
\ee
Using this result and the fact that the Aretakis conserved quantities are non-zero for generic initial data,\footnote{In fact, as shown in~\cite{Lucietti:2012xr}, even for initial data corresponding to vanishing Aretakis conserved quantities, such as an initial ingoing wavepacket, the instability still sets in, albeit at higher radial derivatives.} we will now re-derive the characteristic Aretakis instability.

To begin with, the expression of the $\ell$'th set of Aretakis conservation laws in terms of the near-horizon transverse derivatives of the field, Eq.~\eqref{eq:AretakisNHlimL}, implies that
\be
	A_{\ell m} = \frac{M^{\ell-1}}{\left(\ell+1\right)!}\lim_{r\rightarrow M}\partial_{r}^{\ell}\left[r\partial_{r}\left(r\Phi_{\ell m}\right)\right] \xrightarrow{\upsilon\rightarrow\infty} \frac{M^{\ell+1}}{\left(\ell+1\right)!}\left(\partial_{r}^{\ell+1}\Phi_{\ell m}\right)\big|_{r=M} \,.
\ee
Indeed, the neglected terms that enter the expression of $A_{\ell m}$ in Eq.~\eqref{eq:AretakisNHlimL} are built from a linear combination of $\left(\partial_{r}^{\ell}\Phi_{\ell m}\right)\big|_{r=M}$ and $\left(\partial_{r}^{\ell-1}\Phi_{\ell m}\right)\big|_{r=M}$ which do decay at late times according to Eq.~\eqref{eq:ERNkDrDecay}. This brings us to the first result that $\left(\partial_{r}^{\ell+1}\Phi_{\ell m}\right)\big|_{r=M}$ does \textit{not} decay at late times, namely,
\be\label{eq:ERNAretakisLateTimesL}
	\left(\partial_{r}^{\ell+1}\Phi_{\ell m}\right)\big|_{r=M} \xrightarrow{\upsilon\rightarrow\infty} \frac{\left(\ell+1\right)!}{M^{\ell+1}}A_{\ell m} \,.
\ee
Next, let us slightly rearrange Eq.~\eqref{eq:DkERNWaveEq} into
\be\ba\label{eq:ERNEOMk}
	\partial_{\upsilon}\left(\partial_{r}^{k+1}\Phi_{\ell m}\right)\big|_{r=M} = \frac{1}{M^2}&\bigg\{ \frac{1}{2}\left(\ell-k\right)\left(\ell+k+1\right)\left(\partial_{r}^{k}\Phi_{\ell m}\right)\big|_{r=M} \\
	&\quad-\left(2k+1\right)M\partial_{\upsilon}\left(\partial_{r}^{k}\Phi_{\ell m}\right)\big|_{r=M} \\
	&\quad-k^2\partial_{\upsilon}\left(\partial_{r}^{k-1}\Phi_{\ell m}\right)\big|_{r=M} \bigg\} \,.
\ea\ee
Setting $k=\ell+1$ and taking the late-time behavior, the decaying conditions of Eq.~\eqref{eq:ERNkDrDecay} imply that the RHS of Eq.~\eqref{eq:ERNEOMk} is dominated by a single term, namely,
\be\ba
	\partial_{\upsilon}\left(\partial_{r}^{\ell+2}\Phi_{\ell m}\right)\big|_{r=M} &\xrightarrow{\upsilon\rightarrow\infty} -\frac{\left(\ell+1\right)}{M^2}\left(\partial_{r}^{\ell+1}\Phi_{\ell m}\right)\big|_{r=M} = -\left(\ell+1\right)\left(\ell+1\right)!\frac{A_{\ell m}}{M^{\ell+3}} \,,
\ea\ee
where we used Eq.~\eqref{eq:ERNAretakisLateTimesL}. This can be immediately integrated to show that, at the horizon, $\partial_{r}^{\ell+2}\Phi_{\ell m}$ grows linearly in time as $ \upsilon\rightarrow\infty$,
\be\ba
	\left(\partial_{r}^{\ell+2}\Phi_{\ell m}\right)\big|_{r=M} \xrightarrow{\upsilon\rightarrow\infty} &-\left(\ell+1\right)\left(\ell+1\right)!\frac{A_{\ell m}}{M^{\ell+2}}\frac{\upsilon}{M} \,.
\ea\ee

This growth propagates recursively to higher transverse derivatives, as one can see by taking increasing values of $k$ in Eq.~\eqref{eq:ERNEOMk}. For instance, the next step is setting $k=\ell+2$ and taking the $\upsilon\rightarrow\infty$ limit. Using the just-derived linear blow-up of $\left(\partial_{r}^{\ell+2}\Phi_{\ell m}\right)\big|_{r=M}$ and the further observation that $\partial_{\upsilon}\left(\partial_{r}^{\ell+2}\Phi_{\ell m}\right)\big|_{r=M}$ is subdominant to $\left(\partial_{r}^{\ell+2}\Phi_{\ell m}\right)\big|_{r=M}$ at late times, we obtain that
\be\ba
	\partial_{\upsilon}\left(\partial_{r}^{\ell+3}\Phi_{\ell m}\right)\big|_{r=M} &\xrightarrow{\upsilon\rightarrow\infty} -\frac{\left(2\ell+3\right)}{M^2}\left(\partial_{r}^{\ell+2}\Phi_{\ell m}\right)\big|_{r=M} \propto A_{\ell m} \upsilon,
\ea\ee
from which we infer that $\left(\partial_{r}^{\ell+3}\Phi_{\ell m}\right)\big|_{r=M}$ blows up quadratically as $\upsilon\rightarrow\infty$, explicitly
\be\ba
	\left(\partial_{r}^{\ell+3}\Phi_{\ell m}\right)\big|_{r=M} \xrightarrow{\upsilon\rightarrow\infty} &\frac{1}{2}\left(\ell+1\right)\left(2\ell+3\right)\left(\ell+1\right)!\frac{A_{\ell m}}{M^{\ell+3}}\left(\frac{\upsilon}{M}\right)^2 \,.
\ea\ee
Iterating this procedure, one realizes that
\be\label{eq:ERNRecAdS}
	\partial_{\upsilon}\left(\partial_{r}^{\ell+n+2}\Phi_{\ell m}\right)\big|_{r=M} \xrightarrow{\upsilon\rightarrow\infty} -\frac{1}{2M^2}\left(n+1\right)\left(2\ell+n+2\right)\left(\partial_{r}^{\ell+n+1}\Phi_{\ell m}\right)\big|_{r=M} \,,
\ee
which can be easily solved recursively starting from Eq.~\eqref{eq:ERNAretakisLateTimesL} to give the final result (see Theorem 6 of~\cite{Aretakis:2011hc}) that, \textit{if $A_{\ell m}\ne0$,\footnote{The non-vanishing of the Aretakis conserved quantities was shown to hold for generic initial data in Proposition 7.2.1 in \cite{Aretakis:2011hc}.} then $\left(\partial_{r}^{\ell+k+1}\Phi_{\ell m}\right)\big|_{r=M}$ blows up like $\upsilon^{k}$ as $\upsilon\rightarrow\infty$}, with the explicit behavior being
\begin{empheq}[box=\tcbhighmath]{align}\label{eq:ERNAretakisInstability}
	\left(\partial_{r}^{\ell+k+1}\Phi_{\ell m}\right)\big|_{r=M} \xrightarrow{\upsilon\rightarrow\infty} &\frac{\left(2\ell+k+1\right)!}{\left(2\ell+1\right)!}\left(\ell+1\right)!\frac{A_{\ell m}}{M^{\ell+k+1}}\left(-\frac{\upsilon}{2M}\right)^{k}\,. 
\end{empheq}

Even at the level of linear perturbations on a fixed background, the Aretakis mechanism shows that arbitrarily small initial data can lead to unbounded growth of transverse derivatives along the horizon. Although the field itself and its associated stress-energy tensor $T_{\mu\nu}$ (which contains only first-order derivatives of the field) remain bounded along the horizon, higher transverse derivatives of the field and thus of $T_{\mu\nu}$ exhibit polynomial growth, which reflects a loss a regularity and signals the presence of an instability. Nonlinear effects and backreaction are required to determine the final state of the geometry; it was argued in~\cite{Murata:2013daa} that, under generic perturbations, the endpoint of the instability is a non-extreme RN solution.\footnote{See also \cite{Hadar:2018izi} for an analytic study of backreaction effects for near-extremal black holes using the Jackiw-Teitelboim model for the near-horizon geometry.}

At this point it is interesting to notice that this analytic result is exactly reproduced if one just studied the wave equation in the near-horizon extremal throat geometry, i.e. in the $\text{AdS}_2\times\mathbb{S}^2$ background of~\cite{Bardeen:1999px,Amsel:2009et}
\be
	ds_{\text{NHE}}^2 = -\frac{\left(r-M\right)^2}{M^2}\dd\upsilon^2 + 2\dd\upsilon \dd r + M^2\dd\Omega_2^2 \,,
\ee
for which the wave equation for the spherical harmonic modes of the scalar field perturbation reads
\be
	\left[\partial_{r}\left(r-M\right)^{2}\partial_{r} +2M^2\partial_{\upsilon}\partial_{r} -\ell\left(\ell+1\right)\right]\Phi_{\ell m} = 0 \,.
\ee
In fact, this is identical to the wave equation in the $2$-dimensional pure AdS$_2$ background, $ds_{\text{AdS}_2}^2 = -\frac{\left(r-M\right)^2}{M^2}\dd\upsilon^2 + 2\dd\upsilon \dd r$, for a massive scalar field of (dimensionless) mass-squared $\mu^2=\ell\left(\ell+1\right)$~\cite{Lucietti:2012xr}. Acting with $k$ radial derivatives and taking the near-horizon limit
\be\label{eq:DkAdS2WaveEq}
	\partial_{\upsilon}\left(\partial_{r}^{k+1}\Phi_{\ell m}\right)\big|_{r=M} = \frac{1}{2M^2}\left(\ell-k\right)\left(\ell+k+1\right)\left(\partial_{r}^{k}\Phi_{\ell m}\right)\big|_{r=M} \,,
\ee
reveals that:
\begin{itemize}
	\item There is a hierarchy of conservation laws in the near-horizon throat. Specifically, setting $k=\ell$ in Eq.~\eqref{eq:DkAdS2WaveEq}, shows that $\left(\partial_{r}^{k+1}\Phi_{\ell m}\right)\big|_{r=M}$ is conserved there. Comparing with the full Aretakis conserved quantities $A_{\ell m}$ of Eq.~\eqref{eq:AretakisNHlimL}, this agrees with the expression of $A_{\ell m}$ in terms of the late-time behavior of the scalar perturbation of ERN.
	\item Setting $k=\ell+n+1$ in Eq.~\eqref{eq:DkAdS2WaveEq} exactly reproduces the recursion relation of Eq.~\eqref{eq:ERNRecAdS} from which to infer the Aretakis instability starting from the non-decay of $\left(\partial_{r}^{k+1}\Phi_{\ell m}\right)\big|_{r=M}$ at late times.
\end{itemize}
We may therefore interpret the Aretakis instability of ERN under massless, minimally coupled scalar field perturbations as a purely local effect, arising entirely from the details of the near-horizon throat geometry, with no regards of what happens in the far-horizon region at all. This observation may not come as a surprise, since the near-horizon throat geometry is a ``near-horizon \textit{and} late-time'' region of the full spacetime~\cite{Bardeen:1999px,Amsel:2009et} and is therefore expected to contain information about the Aretakis instability. In particular, as we will review in Section~\ref{sec:ScalinfArgumentConventional}, the Aretakis instability can be understood as a phenomenon arising from the emergent scale invariance in the throat geometry~\cite{Gralla:2016jfc,Gralla:2017lto,Gralla:2018xzo}.

\subsection{Universality of Aretakis instability}

So far we have demonstrated the existence of the Aretakis conserved quantities and the resulting Aretakis instability associated with a (massless, minimally coupled) scalar field that perturbs a four-dimensional ERN black hole. This phenomenon, however, is much more general and, as we will argue here, is a purely local effect related to the near-horizon geometry. For this purpose, let us consider the most general static and spherically symmetric geometry in four spacetime dimensions\footnote{The arguments of this section also apply to black holes in higher spacetime dimensions, with minor modifications to the various expressions coming from employing the spherical harmonics modes of the higher-dimensional sphere.}, equipped with a degenerate black hole horizon located at some finite radial distance. In horizon-centered advanced null Gaussian coordinates $\left(\upsilon,\rho,x^{A}\right)$, the line element can always be brought to the form
\be\label{eq:MetricHorExtremalRank2}
	ds^2 = -\rho^2 b(\rho)\dd\upsilon^2 +2\dd\upsilon \dd\rho +r^2(\rho)\dd\Omega_2^2 \,,
\ee
where $\rho=0$ is the location of the horizon, with areal radius $r(\rho=0)\coloneqq r_{\text{h}}$, and $b(\rho)$ is a theory-dependent function that is smooth near $\rho=0$. In this section, we will focus to conventional extremal horizons, with rank of degeneracy $2$, for which $b(0)\ne0$.

\subsubsection{Conserved quantities and instabilities in the $\ell=0$ sector}
We now consider massless scalar perturbations $\Phi$, minimally coupled to this generic background geometry. After expanding the perturbation into spherical harmonic modes $\Phi_{\ell m}$ on $\mathbb{S}^2$, the wave equation governing the linearized dynamics can be rearranged into,
\be\label{eq:boxGenSphSymm}
	\partial_{\upsilon}\left(r\partial_{\rho}\left(r\Phi_{\ell m}\right)\right) = \frac{1}{2}\left[\ell\left(\ell+1\right)-\partial_{\rho}\left(\rho^2r^2 b\,\partial_{\rho}\right)\right]\Phi_{\ell m} \,.
\ee
Zooming-in on the horizon, it is then clear that the $s$-wave mode gives rise to the first Aretakis conserved quantity,
\be\label{eq:AretakisConstantN2}
	A_{00} \coloneqq \frac{1}{r_{\text{h}}}\lim_{\rho\rightarrow0}\left[r\partial_{\rho}\left(r\Phi_{00}\right)\right] \quad\Rightarrow\quad \partial_{\upsilon}A_{00} = 0 \,. 
\ee
Assuming that $\Phi_{00}$ decays at large advanced times, we then obtain
\be
	\left(\partial_{\rho}\Phi_{00}\right)\big|_{\rho=0}\xrightarrow{\upsilon\rightarrow\infty} \frac{A_{00} }{r_{\text{h}}} \,,
\ee
which shows that the first transverse derivative of the $s$-wave mode does \textit{not} decay on the horizon. Let us now follow the analogous procedure to what we did in the ERN example to reveal an Aretakis instability for this $s$-wave mode. The crucial observation is that, in the late-time regime, $\Phi_{00}$ and $\partial_{\upsilon}\partial_{\rho}\Phi_{00}$ decay and, in particular, are subleading with respect to $\partial_{\rho}\Phi_{00}$. As a result, acting with $\partial_{\rho}$ on Eq.~\eqref{eq:boxGenSphSymm}, taking the $\rho\rightarrow0$ and $\upsilon\rightarrow\infty$ limits and setting $\ell=0$ reveals that
\be
	\partial_{\upsilon}\left(\partial_{\rho}^2\Phi_{00}\right)\big|_{\rho=0} \xrightarrow{\upsilon\rightarrow\infty} -\left(r^2 b \right)\big|_{\rho=0}\frac{1}{r_{\text{h}}^2}\left(\partial_{\rho}\Phi_{00}\right)\big|_{\rho=0} \xrightarrow{\upsilon\rightarrow\infty}-\left(r^2 b \right)\big|_{\rho=0}\frac{A_{00}}{r_{\text{h}}^3}
\ee
and, therefore, the second transverse derivative of the $s$-wave mode grows linearly on the horizon at late advanced times,
\be
	\left(\partial_{\rho}^2\Phi_{00}\right)\big|_{\rho=0} \xrightarrow{\upsilon\rightarrow\infty} -\left(r^2 b\right)\big|_{\rho=0}\frac{A_{00}}{r_{\text{h}}^2}\frac{\upsilon}{r_{\text{h}}} \,.
\ee
Continuing this way for higher transverse derivatives, it is realized that $\left(\partial_{\rho}^{k+1}\Phi_{00}\right)\big|_{\rho=0}$ grows like $\upsilon^{k}$ as $\upsilon\rightarrow\infty$. The exact behavior of this blow-up can be found by taking $k$ radial derivatives of Eq.~\eqref{eq:boxGenSphSymm}, zooming-in on the horizon, setting $\ell=0$ and keeping the leading contributions as $\upsilon\rightarrow\infty$, which outputs the following recursion relation
\be
	\partial_{\upsilon}\left(\partial_{\rho}^{k+1}\Phi_{00}\right)\big|_{\rho=0} \xrightarrow{\upsilon\rightarrow\infty} -\left(r^2 b\right)\big|_{\rho=0}\frac{k\left(k+1\right)}{2r_{\text{h}}^2}\left(\partial_{\rho}^{k}\Phi_{00}\right)\big|_{\rho=0} \,.
\ee
Combining with the initial condition $\left(\partial_{\rho}\Phi_{00}\right)\big|_{\rho=0}\xrightarrow{\upsilon\rightarrow\infty}\frac{1}{r_{\text{h}}}A_{00}$, we then find
\be\label{eq:AretakisInstabilityConventional}
	\left(\partial_{\rho}^{k+1}\Phi_{00}\right)\big|_{\rho=0} \xrightarrow{\upsilon\rightarrow\infty} \left(k+1\right)!\frac{A_{00}}{r_{\text{h}}^{k+1}}\left(-\frac{\left(r^2 b\right)\big|_{\rho=0}\upsilon}{2r_{\text{h}}}\right)^{k} \,.
\ee

In summary, we have found that, \textit{if the $s$-wave mode on the horizon $\Phi_{00}\big|_{\rho=0}$ decays at large advanced times and $A_{00}\ne0$, then $\left(\partial_{\rho}\Phi_{00}\right)\big|_{\rho=0}$ does not decay, while $\left(\partial_{\rho}^{k+1}\Phi_{00}\right)\big|_{\rho=0}$ blows up polynomially like $\upsilon^{k}$ as $\upsilon\rightarrow\infty$}, with the explicit behavior given by Eq.~\eqref{eq:AretakisInstabilityConventional}.\footnote{In fact, even when $\Phi_{00}\big|_{\rho=0}$ approaches a constant, instead of decaying, at late advanced times the above proof still outputs the behavior $\left(\partial_{\rho}^{k+1}\Phi_{00}\right)\big|_{\rho=0} = \mathcal{C}\upsilon^{k}$, but with a proportionality coefficient $\mathcal{C}$ that also depends on $\lim_{\upsilon\rightarrow\infty}\Phi_{00}\big|_{\rho=0}$.} More importantly, this is a universal result that holds for all static and spherically symmetric black holes equipped with a degenerate horizon, regardless of what theory supports such geometries, i.e. regardless of what $b(\rho)$ or $r(\rho)$ are.

Let it also be noted that, unlike the case of ERN, massless scalar probe fields in the background of a generic doubly-degenerate horizon do not possess locally defined Aretakis constants for higher spherical harmonic modes with $\ell\ge1$. Indeed, taking $k$ derivatives of Eq.~\eqref{eq:boxGenSphSymm} before setting $\rho=0$ yields
\be
    \partial_\upsilon \partial_\rho^{k}\left( r \partial_\rho \left( r\Phi_{\ell m}\right) \right)=\frac12 \left[\ell(\ell+1)-k(k+1)(r^2 b)\right]\partial_\rho^k\Phi_{\ell m}+\mathcal O({\partial^{i< k}_\rho \Phi_{\ell m}})
\ee
where $\mathcal O({\partial^{i< k}_\rho \Phi_{\ell m}})$ represents a linear combination of terms with less than $k$ transverse derivatives. Even if these terms can be set to zero, the contribution in square brackets does not generically vanish if $r^2 b\notin \mathbb{N}$ term ($r^2 b=1$ in ERN). Nevertheless, the $\ell\ne0$ modes are still expected to exhibit an instability, similar to what happens with massive scalar perturbations in the background of ERN~\cite{Lucietti:2012xr}.

\subsubsection{Instability from scale invariance of the near-horizon throat}
\label{sec:ScalinfArgumentConventional}
It turns out that the Aretakis instability can be derived from simple arguments that exploit the emergent scale invariance in the near-horizon throat, first developed in \cite{Gralla:2016jfc,Gralla:2017lto,Gralla:2018xzo} for extremal black holes, and recently extended to extremal black $p$-branes in \cite{Chen:2025sim}. We will review here this line of reasoning.

First of all, we introduce the following (dimensionless) near-horizon coordinates
\be
	\bar{\rho} = \frac{\rho}{\lambda}\sqrt{b(0)} \,,\quad \bar{\upsilon} = \lambda \upsilon\sqrt{b(0)} \,,
\ee
to chart the line element of Eq.~\eqref{eq:MetricHorExtremalRank2} and take the $\lambda\rightarrow0$ scaling limit~\cite{Bardeen:1999px,Amsel:2009et}. The resulting geometry
\be
	ds^2 = b^2_{\text{AdS}_2}\left[-\bar{\rho}^2\dd\bar{\upsilon}^2+2\dd\bar{\upsilon} \dd\bar{\rho}\right] +r_{\text{h}}^2\dd\Omega_2^2 +\mathcal{O}\left(\lambda\right) \,,
\ee
is an $\text{AdS}_2\times\mathbb{S}^2$ geometry, with the AdS and $\mathbb{S}^2$ radii given by $b_{\text{AdS}_2}\coloneqq1/\sqrt{b(0)}$ and $r_{\text{h}}$ respectively. This geometry is also known as the Bertotti-Robinson metric and is an exact solution to the Einstein-Maxwell field equations in vacuum (with zero cosmological constant). The appearance of an $\text{AdS}_2$ geometry enhances the isometry group from $\mathbb{R}_{t}\times SO(3)$ to $\SL\times SO(3)$, with the $\SL$ Killing vectors given by
\be\ba
	{}&\xi_{+1} = \partial_{\bar{\upsilon}} \,,\quad \xi_{0} = \bar{\upsilon}\,\partial_{\bar{\upsilon}} -\bar{\rho}\,\partial_{\bar{\rho}} \,,\quad \xi_{-1} = \bar{\upsilon}^2\partial_{\bar{\upsilon}} -2\left(\bar{\upsilon}\bar{\rho}+1\right)\partial_{\bar{\rho}} \,, \\
	&\quad\Rightarrow\quad \left[\xi_{m},\xi_{n}\right] = \left(m-n\right)\xi_{m+n} \,.
\ea\ee
In particular, one immediately sees the emergence of the scale invariance generated by $\xi_0$, corresponding to the finite transformation $\left\{\bar{\upsilon}\rightarrow\bar{\upsilon}/c,\,\bar{\rho}\rightarrow c\bar{\rho}\right\}$. Let us emphasize here that the near-horizon limit $\lambda\rightarrow0$ is, in fact, a ``near-horizon, late-time'' limit, as it corresponds to $\rho\rightarrow 0$ \textit{and} $\upsilon\rightarrow\infty$.

To see how this scale invariance captures the Aretakis instability, we now consider a spacetime tensor field $W$ that is smooth everywhere, including at the horizon. As we approach the horizon at late times, i.e.~in the $\lambda\rightarrow0$ limit, this tensor field must behave like
\be
	W(\upsilon,\rho,x^{A}) = \lambda^{p}\left[\bar{W}(\bar{\upsilon},\bar{\rho},x^{A}) +\mathcal{O}\left(\lambda\right)\right] \,,
\ee
for some exponent $p$ (not necessarily positive), with $\bar{W}$ being independent of $\lambda$. As such, $\bar{W}$ is a tensor field in the near-horizon throat. Since the limit $\lambda\to 0$ is invariant under rescalings of $\lambda$, then $\bar W\to c^p \bar W$ if $\lambda\to c\lambda$. Such a scaling is generated by the dilatation Killing vector field, which means that $\bar{W}$ is conformally Lie-dragged along it,
\be
	\mathcal{L}_{\xi_0}\bar{W} = -p\bar{W} \,.
\ee
In other words, $\bar{W}$ is self-similar with weight $-p$. This self-similarity property is sufficient to demonstrate the Aretakis instability. Consider, for instance, the case of a scalar field $W = \Psi = \lambda^{p}\left[\bar{\Psi} +\mathcal{O}\left(\lambda\right)\right]$. The general solution to $\mathcal{L}_{\xi_0}\bar{\Psi}=-p\bar{\Psi}$ is
\be
	\bar{\Psi}(\bar{\upsilon},\bar{\rho},x^{A}) = \bar{\upsilon}^{-p}F(\bar{\upsilon}\bar{\rho},x^{A}) \,,
\ee
which, translated back to the original spacetime field, leads to the following near-horizon, late-time behavior
\be
	\Psi(\upsilon,\rho,x^{A}) = \upsilon^{-p}G(\upsilon\rho,x^{A}) \quad\text{as $\rho\rightarrow0$ and $\upsilon\rightarrow\infty$} \,,
\ee
where we have denoted for economy $G(\upsilon\rho,x^{A})\coloneqq b(0)^{-p/2}F(\upsilon\rho\, b(0),x^{A})$. As a result, sufficiently high transverse derivatives of the scalar field eventually blow up on the horizon at late times
\be
	\lim_{\rho\rightarrow0}\partial_{\rho}^{n}\Psi \xrightarrow{\upsilon\rightarrow\infty} C_{n}(x^{A}) \upsilon^{n-p} \,,
\ee
which is precisely the celebrated Aretakis instability. The coefficients $C_{n}(x^{A})$ contain the would-be Aretakis constants but are not fixed by this scaling argument.

Let us emphasize that the above line of reasoning is entirely off-shell. The only assumption is that the near-horizon throat admits a dilatation Killing vector generating reciprocal rescalings of $\bar{\rho}$ and $\bar{\upsilon}$ or, equivalently, a rescaling of the parameter $\lambda$, and that the field of interest is sufficiently smooth. For instance, the scaling argument is insensitive to the specific equation of motion satisfied by the scalar field $\Psi$: $\Psi$ may represent a massless minimally coupled real scalar field, a massive electrically charged scalar field (for which no locally-defined Aretakis conserved quantities exist, though an Aretakis instability nevertheless persists~\cite{Lucietti:2012xr}) or even one of the Newman-Penrose scalars. The value of the scaling exponent $p$, however, depends sensitively on the type of field, being determined not only by its tensorial structure but also by its late-time behavior. For example, $p=\ell+1$ for $\Phi_{\ell m}$~\cite{Aretakis:2011hc}.


\section{Screening the Aretakis instability with multi-degenerate horizons}
\label{sec:MultiDegenerateBHs}
Asymptotically flat black hole solutions of general relativity are typically classified as either non-extremal, with two distinct horizons, an outer (event) horizon and an inner (Cauchy) horizon, or extremal, characterized by a single doubly-degenerate event horizon. In asymptotically de Sitter spacetimes, the presence of a positive cosmological constant introduces additional horizon structure at finite radius. The generic non-extremal configuration consists of an outermost cosmological horizon, an event horizon, and an innermost Cauchy horizon (together with an additional, unphysical root at negative radius). In this setting as well, extremal limits generically yield horizons that are at most doubly degenerate.

Beyond these conventional scenarios, asymptotically flat black hole solutions with multiple coincident horizons have been constructed in theories that preserve the gravitational sector of general relativity while modifying the matter content, most notably through non-linear electrodynamics~\cite{Nojiri:2017kex,Gao:2017vqv,Gao:2021kvr}. These constructions allow for a controlled tuning of parameters such that more than two horizons coincide, giving rise to multi-degenerate horizons. Recently, such geometries have been proposed as potential stable endpoints of black hole evolution~\cite{DiFilippo:2024spj,Gao:2025plm}, motivating a closer examination of their dynamical properties. Owing to the presence of a multi-degenerate inner horizon, these configurations potentially evade the mass inflation instability, while preserving a non-extremal outer event horizon. An open question is whether they could also be free of the Aretakis instability. 

In this section, we investigate how the Aretakis instability manifests in the presence of a multi-degenerate horizon. A notable feature of these geometries is the progressive vanishing of higher radial derivatives of $g_{\upsilon\upsilon}$ at the horizon, which suggests that the onset of instability is postponed to increasingly higher-order transverse derivatives. As we will show, the Aretakis instability persists for multi-degenerate horizons, but in a softened form: the blow-up rates of horizon derivatives decrease as the degree of degeneracy increases, effectively screening the instability.

A further motivation for this analysis comes from the properties of photon spheres in these spacetimes. Doubly-degenerate horizons are always accompanied by a photon sphere that lies precisely on the degenerate horizon and is manifestly stable. Since the stability of photon spheres has been conjectured to be linked to the presence of the Aretakis instability~\cite{Gralla:2019isj}, and later argued to be insufficient to characterize it~\cite{Franzin:2023slm}, these configurations provide a natural testing ground for this proposal. We therefore also comment on the stability properties of the photon sphere located on the horizon and contrast them with the persistence, albeit softened, of the Aretakis instability.


\subsection{Black holes with multiple horizons and multi-degenerate horizons}

Let us consider a generic static and spherically symmetric geometry that contains a multi-degenerate horizon, whose rank of degeneracy is equal to $N$. Its metric in horizon-centered advanced null Gaussian coordinates $\left(\upsilon,\rho,x^{A}\right)$ is read from the line element
\be\label{eq:MetricHorN}
	ds^2 = -\rho^{N}b(\rho)\dd\upsilon^2 +2\dd\upsilon \dd\rho +r^2(\rho)\dd\Omega_2^2 \,,
\ee
where $b(\rho)$ is a theory-dependent function that is smooth and non-vanishing near the horizon, $b(0)\ne0$, and, as before, we denote the areal radius of the horizon as $r_{\text{h}}\coloneqq r(\rho=0)$.

It is possible that the spacetime contains more horizons at $\rho\ne 0$, and that such horizons have different ranks of degeneracies. However, since our analysis is local, we can focus on a single horizon and consider $\rho$ to be centered around that.


\subsection{Dependence of the Aretakis instability on the rank of horizon degeneracy}
\label{sec:AretakisHorN}

Using the above class of static and spherically symmetric geometries, let us now re-examine the problem of a massless scalar perturbations $\Phi$, following a procedure analogous to that of Section~\ref{sec:AretakisReview}. In the advanced coordinate system, the equation of motion, $\Box\Phi=0$, expanded into spherical harmonic modes $\Phi_{\ell m}$, reduces to
\be\label{eq:boxNdeg}
	\partial_{\upsilon}\left(r\partial_{\rho}(r\Phi_{\ell m})\right) = \frac{1}{2}\left[\ell\left(\ell+1\right)-\partial_{\rho}\left(\rho^{N}r^2 b\,\partial_{\rho}\right)\right]\Phi_{\ell m} \,.
\ee
Taking the near-horizon limit, $\rho\rightarrow0$, we see the first effect of a higher rank of horizon degeneracy: besides the first Aretakis constant, Eq.~\eqref{eq:AretakisConstantN2}, associated with the $s$-wave perturbation mode that all extremal horizons exhibit, a horizon with degeneracy $N$ possesses $N-2$ additional Aretakis constants. These arise from the same $s$-wave mode but are associated with higher radial derivatives of the field component. Specifically, the $N-1$ Aretakis constants coming from the $s$-wave sector are given by
\be\label{eq:AretakisNswave}
	\begin{gathered}
		A_{00}^{\left(n\right)} = \frac{r_{\text{h}}^{n-1}}{\left(n+1\right)!}\lim_{\rho\rightarrow0}\partial_{\rho}^{n}\left[r\partial_{\rho}\left(r\Phi_{00}\right)\right] \\
		\Rightarrow \quad \partial_{\upsilon}A_{00}^{\left(n\right)} = 0 \,,\quad 0\le n\le N-2 \,.
	\end{gathered}
\ee

Let us now examine the implications of the existence of these constants on the stability or instability of the $s$-wave mode. First of all, the field itself does not blow up on the horizon in the far future, as can be seen, for instance, by numerically integrating the equations of motion, a task presented Section~\ref{sec:NumericalResults} (see Figure~\ref{fig:phiNdegL0}). Then, as detailed in Appendix~\ref{app:AretakisMultiDegenerate}, a careful analysis of the equations of motion shows that, if the Aretakis constants $A_{00}^{\left(n\right)}$ are non-zero, then the late time behavior of $\left(\partial_{\rho}^{k}\Phi_{00}\right)\big|_{\rho=0}$ is dictated by
\be
\label{eq:FieldDerHorN}
	\left(\partial_{\rho}^{k}\Phi_{00}\right)\big|_{\rho=0} \xrightarrow{\upsilon\rightarrow\infty} a_{N,k}\left(\frac{\upsilon}{r_{\text{h}}}\right)^{\left\lceil\frac{k}{N-1}\right\rceil-1} \,,
\ee
where $a_{N,k}$ are linear combinations of $A_{00}^{\left(n\le q\right)}$, with $q=k-N-\left(N-1\right)\left\lceil\frac{k}{N-1}\right\rceil$, and $\lim_{\upsilon\rightarrow\infty}\Phi_{00}\big|_{\rho=0}$, whose explicit expression can be found case-by-case by following the prescription of Appendix~\ref{app:AretakisMultiDegenerate}. As we will see in Section~\ref{sec:NumericalResults} this behavior is confirmed by the numerical results.

Summarizing the results of this section, we see that, \textit{if} $\Phi_{00}\big|_{\rho=0}$ does not blow up at late times, then $\left(\partial_{\rho}^{k}\Phi_{00}\right)\big|_{\rho=0}$ does \textit{not} decay: in particular, it approaches a constant for $k\le N-1$ and blows up with a power law $\upsilon^{p-1}$, where $p=\left\lceil\frac{k}{N-1}\right\rceil$, for higher values of $k$. This means that, as $N$ increases, more derivatives tend to constant values and the instability kicks on later, in the sense that the first transverse derivative that blows up (linearly) is the $N$-th one.
This analysis suggests that higher horizon degeneracy implies a softening of the Aretakis instability, with the obvious tentative conclusion that an infinitely-degenerate horizon would have no Aretakis instability at all. We will study this possibility in Section~\ref{sec:InfDegBH}, after we corroborate our result with a different approach.

\paragraph{Aretakis instability vs Photon sphere stability} The Aretakis instability is sometimes linked to the presence of a stable photon sphere on the ($N=2$) degenerate horizon~\cite{Gralla:2019isj}. While the presence of a photon sphere on the horizon persists for higher ranks $N$ of degeneracy as well, the stability analysis of such spherical photon orbits does depend on $N$, as well as the degeneracies of the other outer horizons. More specifically, suppose that the $\rho=0$ horizon has rank of degeneracy $N$ and that it is inside a series of $k$ outer horizons, with the $i$'th horizon having degeneracy rank $N_{i}$. Then, as demonstrated more explicitly in Appendix~\ref{app:PhotonSphereStability}, the photon sphere on the $N$-tuply degeneracy horizon at $\rho=0$ is stable if $N$ is even and $\sum_{i=1}^{k}N_{i}$ is also even, it is unstable if $N$ is even and $\sum_{i=1}^{k}N_{i}$ is odd, while it is metastable if $N$ is odd.

On the other hand, the Aretakis instability is always present for any $N$, albeit entering at higher transverse derivatives as $N$ increases. This confirms that the presence of a photon sphere on the degenerate horizon is not sufficient for characterizing the Aretakis instability~\cite{Franzin:2023slm}, here justified by the fact that the latter emerges for any finite $N\ge2$, while the former is not guaranteed to be stable.


\subsection{Instability from scale invariance of multi-degenerate horizons}

Let us now extend the scaling argument of Sec.~\ref{sec:ScalinfArgumentConventional} to a multi-degenerate horizon, again located at the origin of a static and spherically symmetric geometry,
\be
    ds^2 = -\rho^{N}b(\rho)\dd\upsilon^2 +2\dd\upsilon \dd\rho +r^2(\rho)\dd\Omega_{d-2}^2 \,.
\ee
To find the corresponding near-horizon throat geometry, we need to take an inhomogeneous scaling limiting procedure. Namely, we need to take the $\lambda\rightarrow0$ limit of the following coordinates
\be\label{eq:NHlimitNdeg}
	\bar{\rho} =\frac{1}{\lambda}\frac{\rho}{r_0} \,,\quad \bar{\upsilon} = \lambda^{N-1}\frac{\upsilon}{r_0} \,,
\ee
where the length scale $r_0$ is given by
\be
    r_0 \coloneqq \left[b(0)\right]^{-1/N} \,.
\ee
The resulting geometry is
\be
    ds^2 = \frac{r_0^2}{\left(\lambda \bar{\rho}\right)^{N-2}}\left[-\bar{\rho}^{2\left(N-1\right)}\dd\bar{\upsilon}^2 +2\bar{\rho}^{N-2}\dd\bar{\upsilon}\dd\bar{\rho}\right] +r_{\text{h}}^2\dd\Omega_{d-2}^2 + \dots \,.
\ee
For $N=2$, one recovers the Bertotti-Robinson metric. For $N>2$, however, the spherical factor becomes subleading and one ends up with a $2$-dimensional effective metric that is conformally Lifshitz invariant,
\be
	ds^2\bigg|_{N>2} = \Omega^2(\lambda)\left[ds_{\text{Lifshitz}}^2 +\mathcal{O}\left(\lambda\right)\right] \,,\quad ds_{\text{Lifshitz}}^2 = -\bar{\rho}^{2\left(N-1\right)}\dd\bar{\upsilon}^2 +2\bar{\rho}^{N-2}\dd\bar{\upsilon}\dd\bar{\rho} \,.
\ee
The near-horizon throat again exhibits an enhancement: the $\mathbb{R}_{t}$ isometry subgroup generated by $\partial_{t}$ is enhanced to the $\mathfrak{sl}\left(2,\mathbb{R}\right)$ generated by
\be\ba
	{}&\xi_{+1} = \partial_{\bar{\upsilon}} \,,\quad \xi_0 = \bar{\upsilon}\partial_{\bar{\upsilon}} -\frac{\bar{\rho}}{N-1}\partial_{\bar{\rho}} \,, \\
    &\xi_{-1} = \bar{\upsilon}^2\partial_{\bar{\upsilon}} -\frac{2}{N-1}\left(\bar{\upsilon}\bar{\rho}+\frac{1}{\left(N-1\right)\bar{\rho}^{N-2}}\right)\partial_{\bar{\rho}} \,, \\
    &\quad\Rightarrow\quad \left[\xi_{m},\xi_{n}\right] = \left(m-n\right)\xi_{m+n} \,.
\ea\ee

Now, we take again a tensor field $W$ that behaves like
\be\label{eq:NHexpansion}
	W(\upsilon,\rho,x^A) = \lambda^{p}\left[\bar{W}(\bar\upsilon,\bar\rho,x^A)+\mathcal{O}\left(\lambda\right)\right] \,,
\ee
for some $p$. The tensor field $\bar{W}$, defined in the near-horizon throat, inherits the transformation $\bar W\to c^p \bar W$ when $\lambda\to c\lambda$, hence it satisfies the self-similarity property
\be
	\mathcal{L}_{\xi_0}\bar{W} = -\frac{p}{N-1}\bar{W} \,.
\ee
For a scalar field, $W = \Psi = \lambda^{p}\left[\bar{\Psi}+\mathcal{O}\left(\lambda\right)\right]$, the general solution to the previous equation is
\be
	\bar{\Psi}\left(\bar\upsilon,\bar\rho,x^{A}\right) = \bar{\upsilon}^{-p/\left(N-1\right)}F\left(\bar{\upsilon}^{1/\left(N-1\right)}\bar{\rho},x^{A}\right) \,,
\ee
which implies, for the original spacetime scalar field $\Psi$, the following late-time, near-horizon behavior
\be
	\Psi\left(\upsilon,\rho,x^{A}\right) = \upsilon^{-p/\left(N-1\right)}G\left(\upsilon^{1/\left(N-1\right)}\rho,x^{A}\right) \quad\text{as $\rho\rightarrow0$ and $\upsilon\rightarrow\infty$} \,.
\ee
This reveals the following Aretakis instability for the late time behavior of near-horizon transverse derivatives,
\be
	\lim_{\rho\rightarrow0}\partial_{\rho}^{n} \Psi\left(\upsilon,\rho,x^{A}\right) \xrightarrow{\upsilon\rightarrow\infty} C_{n}(x^{A})\upsilon^{\left(n-p\right)/\left(N-1\right)} \,.
\ee

Therefore, as $N$ increases, the near-horizon blow-up becomes slower and slower in $\upsilon$, for every value of $n>p$, with the expectation that the Aretakis instability is completely eliminated in the $N\rightarrow\infty$ limit, which corresponds to an infinitely-degenerate horizon, to which we soon direct our attention.

\paragraph{What happens to the $\ell\ne0$ modes on the near-horizon throat?} We will now argue that only the $\ell=0$ mode of the scalar field survives on the near-horizon throat. To do this, we take the equation of motion Eq.~\eqref{eq:boxNdeg} and apply the scaling limit of Eq.~\eqref{eq:NHlimitNdeg} that zooms-in on the near-horizon throat to get
\be\ba\label{eq:PhiEOMlambda}
    0&=\partial_\upsilon[r\partial_\rho (r \Phi_{\ell m})]-\frac{1}{2}\ell(\ell+1)\Phi_{\ell m}+\frac{1}{2}\partial_\rho(\rho^N r^2 b\partial_\rho\Phi_{\ell m}) \\
    &= \lambda^{N-2}\frac{1}{\bar\rho}\bqty{\frac{1}{r_0}\partial_{\bar\upsilon}[\bar\rho\partial_{\bar\rho}(r\Phi_{\ell m})]+\frac{1}{2}\bar\rho\partial_{\bar\rho}\pqty{\frac{r^2 b(\rho)}{r_0^2 b(0)} \bar\rho^{N-1} \partial_{\bar\rho}\Phi_{\ell m}}}-\frac{1}{2}\ell(\ell+1)\Phi_{\ell m}\,.
\ea\ee
Thus, for $N>2$, the $\lambda\to 0$ limit kills the first set of terms in square brackets, assuming that $\Phi_{\ell m}$ is smooth in the near-horizon throat
\be
 	\text{for $N>2$ :}\quad \ell\left(\ell+1\right)\lim_{\lambda\rightarrow0}\Phi_{\ell m} = 0 \,,
\ee
from which we infer that only the $\ell=0$ mode can survive in the near-horizon throat. This persists for the late time behavior of all the transverse derivatives of the field on the horizon,
\be
    \left(\partial_{\rho}^{k}\Phi_{\ell m}\right)\big|_{\rho=0} \xrightarrow{\upsilon\rightarrow\infty} 0 \quad\text{for $\ell\ne0$} \,,
\ee
as can be checked explicitly by acting with $\partial_{\rho}^{k}$ onto Eq.~\eqref{eq:PhiEOMlambda} and expanding in $\lambda$. As a result, the Aretakis instability for horizons with rank of degeneracy $N>2$ only manifests in the $s$-wave sector, while all $\ell\ne0$ modes decay on the horizon at late times. The absence of higher angular modes is consistent with the metric becoming degenerate in the angular directions for $N>2$.

\subsection{Infinitely-degenerate horizons as candidates of classically stable geometries}
\label{sec:InfDegBH}

As already hinted in this section, a horizon whose rank of degeneracy is infinite appears to be an Aretakis-stable candidate. Indeed, as the rank $N$ of degeneracy becomes larger, the Aretakis instability kicks-in at a higher transverse derivative of the probe field. Here we will study one such class of spacetimes equipped with an infinitely-degenerate horizon and show that no transverse derivatives of the probe field blow up at late times on the horizon.

The geometries we will consider are of the following form, in horizon-centered advanced null Gaussian coordinates,
\be\label{eq:InfDegMetric}
	ds^2=-e^{-r_0^2/\rho^2}b(\rho) \dd\upsilon^2+2 \dd\upsilon \dd\rho +r^2(\rho) \dd\Omega^2_2 \,,
\ee
with $b(\rho)$ again an arbitrary function that is analytic near the horizon and non-vanishing there, $b(0)\ne0$. This metric is smooth everywhere but non-analytic near the horizon, since all the derivatives of $g_{\upsilon\upsilon}$ vanish there, reflecting the fact that the horizon at $\rho=0$ is infinitely degenerate. Nevertheless, there is no curvature singularity at $\rho=0$, since all curvature invariants are regular near the horizon
\be\ba
	R &= \frac{2}{r_{\rm h}^2}+\order{e^{-r_0^2/\rho^2}} \,,\\
	R_{\mu\nu}R^{\mu\nu} &= \frac{2}{r_{\rm h}^4}+\order{e^{-r_0^2/\rho^2}} \,,\\
	R_{\mu\nu\kappa\sigma}R^{\mu\nu\kappa\sigma} &= \frac{4}{r_{\rm h}^4}+\order{e^{-r_0^2/\rho^2}} \,.
\ea\ee

We claim that the infinitely-degenerate horizon of the geometry of Eq.~\eqref{eq:InfDegMetric} is an example of an Aretakis-stable configuration. We will support this claim by examining the near-horizon late-time behavior of a minimally coupled and massless scalar probe field $\Phi$, whose equation of motion, $\Box\Phi=0$, separates on each spherical harmonic mode $\Phi_{\ell m}$ as
\be\label{eq:boxInfDeg}
	\partial_{\upsilon}\left(r\partial_{\rho}\left(r\Phi_{\ell m}\right)\right) = \frac{1}{2}\left[\ell\left(\ell+1\right)-\partial_{\rho}\left(e^{-r_0^2/\rho^2}r^2b\,\partial_{\rho}\right)\right]\Phi_{\ell m}
\ee

\paragraph{Absence of Aretakis instability for $s$-wave modes} In Section~\ref{sec:MultiDegenerateBHs}, we noticed that only the $s$-wave mode of perturbations of the multi-degenerate horizon was prone to the Aretakis instability. As we will see right away, this instability disappears for the infinitely-degenerate horizons of interest. The crucial observation is that the $\ell=0$ sector of Eq.~\eqref{eq:boxInfDeg} is equipped with infinitely-many Aretakis constants, thanks to the infinite degeneracy of the horizon, namely,
\be\label{eq:AretakisConstantsInfDeg}
	\begin{gathered}
		A_{00}^{\left(n\right)} = \frac{r_{\text{h}}^{n-1}}{\left(n+1\right)!}\lim_{\rho\rightarrow0}\partial_{\rho}^{n}\left[r\partial_{\rho}\left(r\Phi_{00}\right)\right] \\
		\Rightarrow \quad \partial_{\upsilon}A_{00}^{\left(n\right)} = 0 \,,\quad \forall n\in\mathbb{N}_0 \,.
	\end{gathered}
\ee
More explicitly, $A_{00}^{\left(k\right)}$ above is a linear combination of $\left(\partial_{\rho}^{n\le k+1}\Phi_{00}\right)\big|_{\rho=0}$. In particular, as explained in Appendix~\ref{app:AretakisMultiDegenerate}, this allows to write $\left(\partial_{\rho}^{k\ge1}\Phi_{00}\right)\big|_{\rho=0}$ as a linear combination of $\Phi_{00}\big|_{\rho=0}$ and $A_{00}^{\left(n\le k-1\right)}$,
\be
	\left(\partial_{\rho}^{k}\Phi_{00}\right)\big|_{\rho=0} = \sum_{n=0}^{k-1}c_{k,n}A_{00}^{\left(n\right)} +d_{k}\Phi_{00}\big|_{\rho=0} \,,
\ee
with $c_{k,n}$ and $d_{k}$ numerical coefficients built from the near-horizon Taylor coefficients of $r\left(\rho\right)$, whose explicit expressions are not important here. As a result, if $\Phi_{00}\big|_{\rho=0}$ does not blow up at late advanced times, then the near-horizon transverse derivatives of the scalar field approach a constant controlled by the Aretakis constants and $\lim_{\upsilon\rightarrow\infty}\Phi_{00}\big|_{\rho=0}$,
\be\ba
	{}&\text{If $\Phi_{00}\big|_{\rho=0}\xrightarrow{\upsilon\rightarrow\infty}\text{const.}$} \,, \\
	&\text{then $\left(\partial_{\rho}^{k}\Phi_{00}\right)\big|_{\rho=0} \xrightarrow{\upsilon\rightarrow\infty} \sum_{n=0}^{k-1}c_{k,n}A_{00}^{\left(n\right)} +d_{k}\lim_{\upsilon\rightarrow\infty}\Phi_{00}\big|_{\rho=0}= \text{const.}$}
\ea\ee
More importantly, this is true for all integer $k\ge1$, thanks to the existence of infinitely-many Aretakis constants. For comparison, the analogous statement for the case of an $N$-degenerate horizon was true for $1\le k\le N-1$.

Summarizing, we have just demonstrated that the $s$-wave mode of a scalar probe field in the background of a static and spherically symmetric spacetime of the form of Eq.~\eqref{eq:InfDegMetric}, equipped with an infinitely-degenerate horizon at $\rho=0$, does not suffer from an Aretakis instability, i.e. none of its transverse derivatives blow up on the horizon at late advanced times. In fact, this absence of Aretakis instability for the $s$-wave mode remains valid for \textit{any} infinitely-degenerate horizon. Indeed, if one considers a generic geometry of the form $ds^2 = -f\left(\rho\right)\dd\upsilon^2+2\dd\upsilon\dd\rho+r^2\left(\rho\right)\dd\Omega_2^2$ equipped with an infinitely-degenerate horizon at $\rho=0$, that is, with $\left(\partial_{\rho}^{k}f\right)\big|_{\rho=0}=0$ for all $k\in\mathbb{N}_0$, then one still finds the infinitely-many Aretakis constants of Eq.~\eqref{eq:AretakisConstantsInfDeg} from which to infer that none of the transverse derivatives $\left(\partial_{\rho}^{k}\Phi_{00}\right)\big|_{\rho=0}$ blow up at late times, if $\lim_{\upsilon\rightarrow\infty}\Phi_{00}\big|_{\rho=0}$ is finite.

As we will see in Section~\ref{sec:NumericalResults}, a numerical analysis resolves two subtleties not addressed by the above arguments; namely it confirms (1) the stability for initial data for which $A_{00}^{\left(n\right)}=0$, e.g. an initial ingoing wavepacket, and (2) the absence of potential non-linear effects at finite time instants, arising from an uncontrolled growth of $\left(\partial_{\rho}^{k}\Phi_{00}\right)\big|_{\rho=0}$ at finite values of $\upsilon$.

\section{Numerical analysis}
\label{sec:NumericalResults}

So far, we have been exploring near-horizon properties of geometries equipped with multi-degenerate horizons using analytical methods that are based on local arguments. We will now switch to a numerical study of Eq.~\eqref{eq:boxNdeg} and Eq.~\eqref{eq:boxInfDeg}.

To study the field dynamics near the horizon, we will formulate the evolution as a characteristic initial value problem, prescribing initial data on intersecting null hypersurfaces, rather than a spacelike Cauchy surface, thus exploiting the causal structure of the spacetime to integrate the equations of motion directly along light-like trajectories.

\paragraph{Construction of double null coordinates regular at the horizon} The first step in achieving this is the construction of double null coordinates that are regular on the horizon. For a generic static and spherically symmetric geometry, the line element in horizon-centered null Gaussian coordinates $\left(\upsilon,\rho,x^{A}\right)$ is given by
\be
    ds^2 = -f\left(\rho\right)\,\dd\upsilon^2+2\dd\upsilon\dd\rho+r^2\left(\rho\right)\dd\Omega_2^2 \,.
\ee
The conventional double null coordinates, $\left(\upsilon,u,x^{A}\right)$, can be obtained by introducing the retarded time according to
\be\label{eq:uRet}
	u = \upsilon -2r_{\ast}\left(\rho\right) \,,
\ee
where $r_{\ast}$ is the tortoise coordinate mapping, given by
\be
	r_{\ast}\left(\rho\right) = \int^{\rho}\frac{d\rho^{\prime}}{f\left(\rho^{\prime}\right)} \,,
\ee
up to a choice of the integration constant. In these coordinates, the geometry reads
\be
	ds^2 = -f\left(\rho\right)\dd\upsilon \dd u +r^2(\rho)\dd\Omega_2^2 \,,
\ee
with $\rho$ implicitly defined by Eq.~\eqref{eq:uRet}. To make this coordinate system regular on the horizon, we follow~\cite{Lucietti:2012xr} and introduce the coordinate $U$ through the tortoise mapping according to
\be
	u = \upsilon_0-2r_{\ast}(-U) \,,
\ee
such that $\frac{du}{dU} = \frac{2}{f\left(-U\right)}$ and the horizon of interest is located at $U=0$, where $u\rightarrow+\infty$, with the exterior of the black hole corresponding to $U<0$. The arbitrary integration constant $\upsilon_0$ is introduced for later convenience and will be fixed to the initial advanced time corresponding to one of the intersecting null hypersurfaces on which the initial data are specified. Then, the line element reads
\be\label{eq:MetricGenericUu}
	ds^2 = -2\frac{f(\rho)}{f(-U)}\dd\upsilon\dd U +r^2(\rho)\dd\Omega_2^2 \,,
\ee
and is now regular at $U=0$.

The relation between $\rho$ and $\left(\upsilon,U\right)$ is found by solving
\be\label{eq:rhoUv0}
	r_{\ast}\left(\rho\left(\upsilon,U\right)\right) = \frac{\upsilon-u}{2} = \frac{\upsilon-\upsilon_0}{2} +r_{\ast}\left(-U\right) \,.
\ee
In particular, at the initial advanced time instant
\be
    \rho\left(\upsilon_0,U\right)=-U \,,
\ee
and, therefore,
\be
    \partial_{\rho}^{n}\big|_{\upsilon=\upsilon_0} = \left(-1\right)^{n}\partial_{U}^{n} \,,
\ee
a relation that will prove useful in extracting the initial conditions needed to evolve the near-horizon transverse derivatives. In fact, as detailed in Appendix~\ref{app:rhoUv}, one can write down the following explicit asymptotic expansion near the origin
\be\label{eq:rhoUv1}
    \rho\left(\upsilon,U\right) \sim -U+\sum_{n=1}^{\infty}\left(\frac{\upsilon-\upsilon_0}{2r_{\text{h}}}f\left(-U\right)\right)^{n}\mathcal{P}_{n}\left(F_1\left(-U\right),F_2\left(-U\right),\dots,F_{n-1}\left(-U\right)\right) \,,
\ee
where $\mathcal{P}_{n}\left(x_1,x_2,\dots,x_{n-1}\right)$ is a polynomial which can be constructed systematically using Eq.~\eqref{eq:PnBnk}, and
\be\label{eq:rhoUv2}
    F_{n}\left(\rho\right) \coloneqq \frac{r_{\text{h}}^{n}}{n!}f\left(\rho\right)\frac{d^{n}}{d\rho^{n}}\frac{1}{f\left(\rho\right)}  \,.
\ee
The first few terms in the expansion are written explicitly in Eq.~\eqref{eq:rhoUv4}.

Numerically, however, the relation between $\rho$ and $\left(\upsilon,U\right)$ is extracted using the following trick: The defining equation, Eq.~\eqref{eq:rhoUv0}, can be recast into the first-order differential equation $\partial_{\upsilon}\rho=\frac{f\left(\rho\right)}{2}$, with initial condition $\rho\left(\upsilon_0,U\right)=-U$, the crucial observation being that this initial condition is exact and not just an approximation that is valid only near the $U$-origin. Then, for each value of $U$, one can numerically integrate the resulting one-dimensional ODE to find the evolved function $\rho\left(\upsilon\ge\upsilon_0,U\right)$.

\paragraph{Equation of motion and initial data} We now move on to writing down the equation we wish to numerically integrate, i.e. the equation of motion for a probe massless scalar field $\Phi$ in the background of Eq.~\eqref{eq:MetricGenericUu}, $\Box\Phi=0$. More specifically, we are interested in the radiative spherical harmonic modes
\be
	\Phi\left(\upsilon,U,x^{A}\right) = \sum_{\ell,m}Y_{\ell m}\left(x^{A}\right)\frac{\varphi_{\ell m}\left(\upsilon,U\right)}{r\left(\rho\right)} \,,
\ee
for which the equation of motion reduces to
\be\label{eq:ODEDoubleNull1}
	-4\partial_{\upsilon}\partial_{U}\varphi_{\ell m} = \mathcal{V}_{\ell}\left(\upsilon,U\right)\varphi_{\ell m} \,,
\ee
the effective potential being given by
\be\label{eq:ODEDoubleNull2}
	\mathcal{V}_{\ell}\left(\upsilon,U\right) = \frac{2f\left(\rho\right)}{f\left(-U\right)}\left\{\frac{\ell\left(\ell+1\right)}{r^2\left(\rho\right)} +\frac{1}{r\left(\rho\right)}\frac{d}{d\rho}\left(f\left(\rho\right)r^{\prime}\left(\rho\right)\right)\right\} \,,
\ee
and where it is understood that $\rho=\rho\left(\upsilon,U\right)$ as per Eq.~\eqref{eq:rhoUv0}.

Following~\cite{Lucietti:2012xr}, initial data will be prescribed on the following null surfaces, intersecting at $\left(\upsilon_0,U_0\right)$,
\be
    \Sigma_0 = \left\{\upsilon=\upsilon_0,U\ge U_0\right\}\cup\left\{\upsilon\ge\upsilon_0,U=U_0\right\} \,.
\ee
namely we will consider either an outgoing wavepacket,
\be
    \varphi_{\ell m}\left(\upsilon_0,U\right) = e^{-\left(U-\mu\right)^2/(2\sigma^2)} \,,\quad \varphi_{\ell m}\left(\upsilon,U_0\right) = 0 \,,
\ee
or an ingoing wavepacket,
\be
    \varphi_{\ell m}\left(\upsilon_0,U\right) = 0 \,,\quad \varphi_{\ell m}\left(\upsilon,U_0\right) = e^{-\left(\upsilon-\mu^{\prime}\right)^2/(2\sigma^{\prime2})} \,,
\ee
where we are suppressing the orbital and azimuthal indices $\ell$ and $m$ in the parameters $\left(\mu,\sigma\right)$ and $\left(\mu^{\prime},\sigma^{\prime}\right)$.

These two types of initial conditions have the property that their $s$-wave mode has either zero or non-zero Aretakis constants. Specifically, the ingoing wavepacket has vanishing Aretakis constants by construction, since the field is not initially supported on the horizon at $U=0$, while the outgoing wavepacket has $A_{00}^{\left(n\right)}\ne0$.

\subsection{Multiply-degenerate horizons}

So far, we have been setting up the problem for a generic static and spherically symmetric background geometry, i.e. without specifying the metric function $f\left(\rho\right)$ or the areal radius $r\left(\rho\right)$. We will now focus to asymptotically flat geometries equipped with a multi-degenerate horizon of rank $N$, with the specific choice
\be
    f\left(\rho\right) = \left(\frac{\rho}{\rho+r_{\text{h}}}\right)^{N} \,.
\ee
We will assume that this on-shell geometry is furthermore such that the areal radius is also an affine parameter, i.e. that $r^{\prime}\left(\rho\right)=1$, and, hence,\footnote{Let it be noted here that, for $r^{\prime}\left(\rho\right)=1$, the non-zero Aretakis constants of Eq.~\eqref{eq:AretakisNswave}, associated with the $s$-wave mode, have the following explicit expression in terms of the parameters $\left(\mu,\sigma\right)$ that characterize the outgoing wavepacket
\begin{equation*}
    A_{00}^{\left(n\right)}=\left(-\frac{r_{\text{h}}}{\sigma}\right)^{n+1}\frac{e^{-\mu^2/(2\sigma^2)}}{r_{\text{h}}\left(n+1\right)!}\left[H_{n+1}\left(\frac{\mu}{\sigma}\right)-n\frac{\sigma}{r_{\text{h}}}H_{n}\left(\frac{\mu}{\sigma}\right)\right] \,,
\end{equation*}
with $H_{n}\left(x\right)$ the $n$'th Hermite polynomial.}
\be
    r\left(\rho\right)=\rho+r_{\text{h}} \,.
\ee

With these choices for the functions, the numerical algorithm for solving the evolution equation, Eqs.~\eqref{eq:ODEDoubleNull1}-\eqref{eq:ODEDoubleNull2}, is presented in Appendix~\ref{app:NumericalSetup}. Following the numerical methods, we generate the behavior of the field and its radial derivatives near the horizon for extremal black holes with degeneracy $N=\{2,3,4\}$ to corroborate the results obtained in Section~\ref{sec:AretakisHorN}. The grid used for this scope was spanned by $\left\{\upsilon_{i},i=0,\dots,N_{\upsilon}\right\}\cup\left\{U_{a},a=0,\dots,N_{U}\right\}$, bounded by $U_0=-0.5$, $U_{N_{U}}=0.0$ and $\upsilon_0=0.0$, $\upsilon_{N_{\upsilon}}=1.0\cdot10^{4}$, with resolution $N_{\upsilon}\times N_{U}=10^{5}\times10^{3}$.

\begin{figure}[t]
    \centering
    \includegraphics[width=1.0\linewidth]{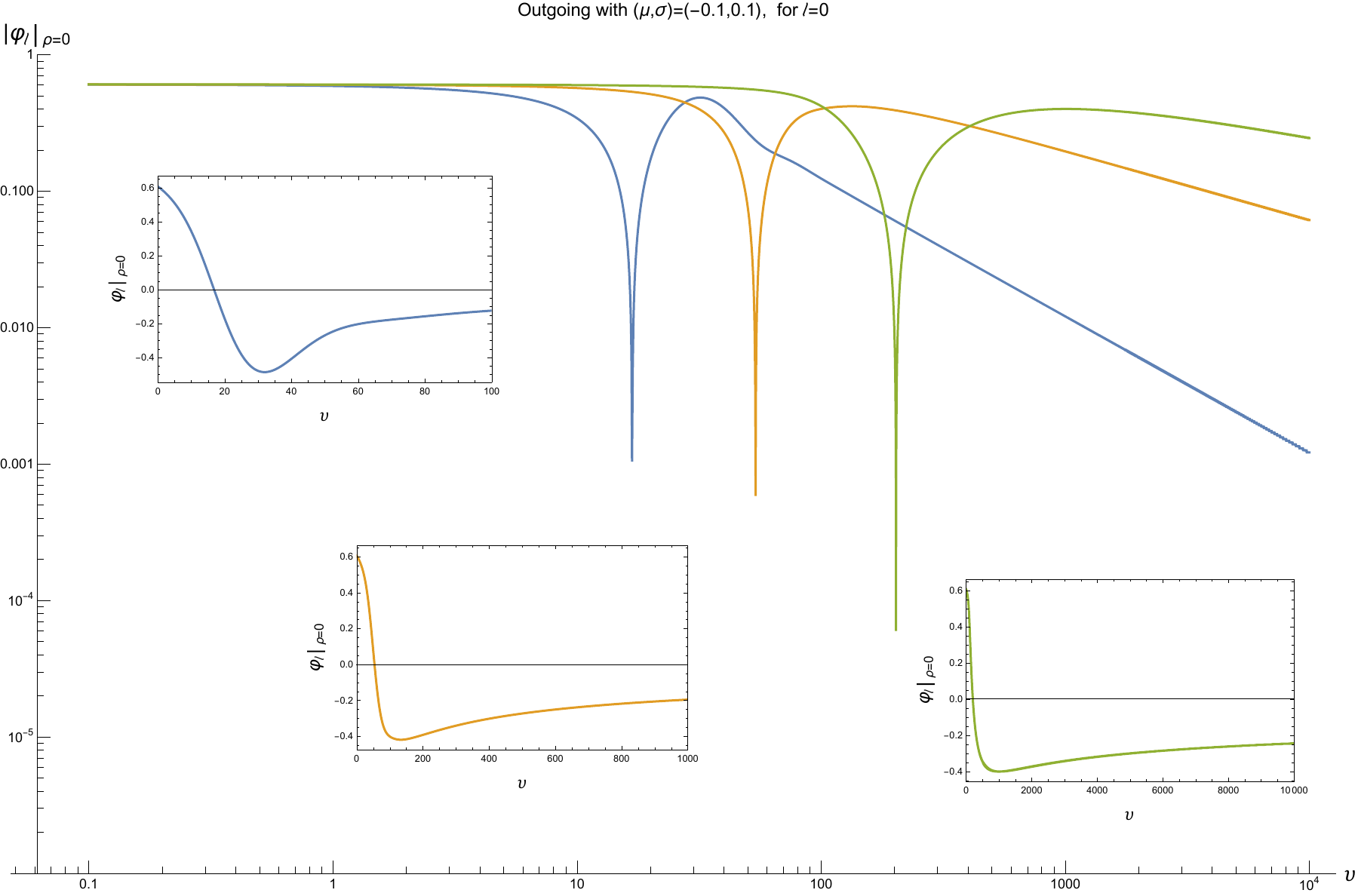}
    \caption{Time evolution of the near-horizon radiative $s$-wave mode $\varphi_{\ell=0}\big|_{\rho=0}$ --- in absolute value in the larger figure --- for ERN ($N=2$, blue) and multi-degenerate horizons with degeneracy ranks $N=3$ (orange) and $N=4$ (green). The initial data correspond to an outgoing wavepacket with $\left(\mu,\sigma\right)=\left(-0.1,0.1\right)$. The inset smaller figures show the same color-coded curves on linear–linear scales, highlighting the distinct characteristic timescales associated with each background configuration.}
    \label{fig:phiNdegL0}
\end{figure}

Starting with the case of initial outgoing condition for the field on the horizon (i.e. non-zero Aretakis constants), with parameters $\left(\mu,\sigma\right)=\left(-0.1,0.1\right)$, we see in Figure~\ref{fig:phiNdegL0} that the $s$-wave ($\ell=0$) mode of the field decays at late time for each of these horizons, albeit the decay becomes slower as the degeneracy is increased. This is agreement with the assumption made in Section~\ref{sec:AretakisHorN} that the field always approaches a constant on the horizon at late advanced times. In fact, fitting $\varphi_{\ell=0}\big|_{\rho=0}$ to a power law decay $\upsilon^{\alpha}$, we find $\alpha=-1.001$, $\alpha=-0.506$ and $\alpha=-0.335$ for $N=2$, $N=3$ and $N=4$ respectively, suggesting that
\be
    \varphi_{\ell=0}\big|_{\rho=0} \propto \upsilon^{-\frac{1}{N-1}} \quad\text{as $\upsilon\rightarrow\infty$} \,.
\ee
These fittings were performed for $\upsilon\ge2000.0$, $\upsilon\ge4000.0$ and $\upsilon\ge9500.0$ for $N=2$, $N=3$ and $N=4$ respectively.

\begin{figure}[t]
    \centering
    \includegraphics[width=1.0\linewidth]{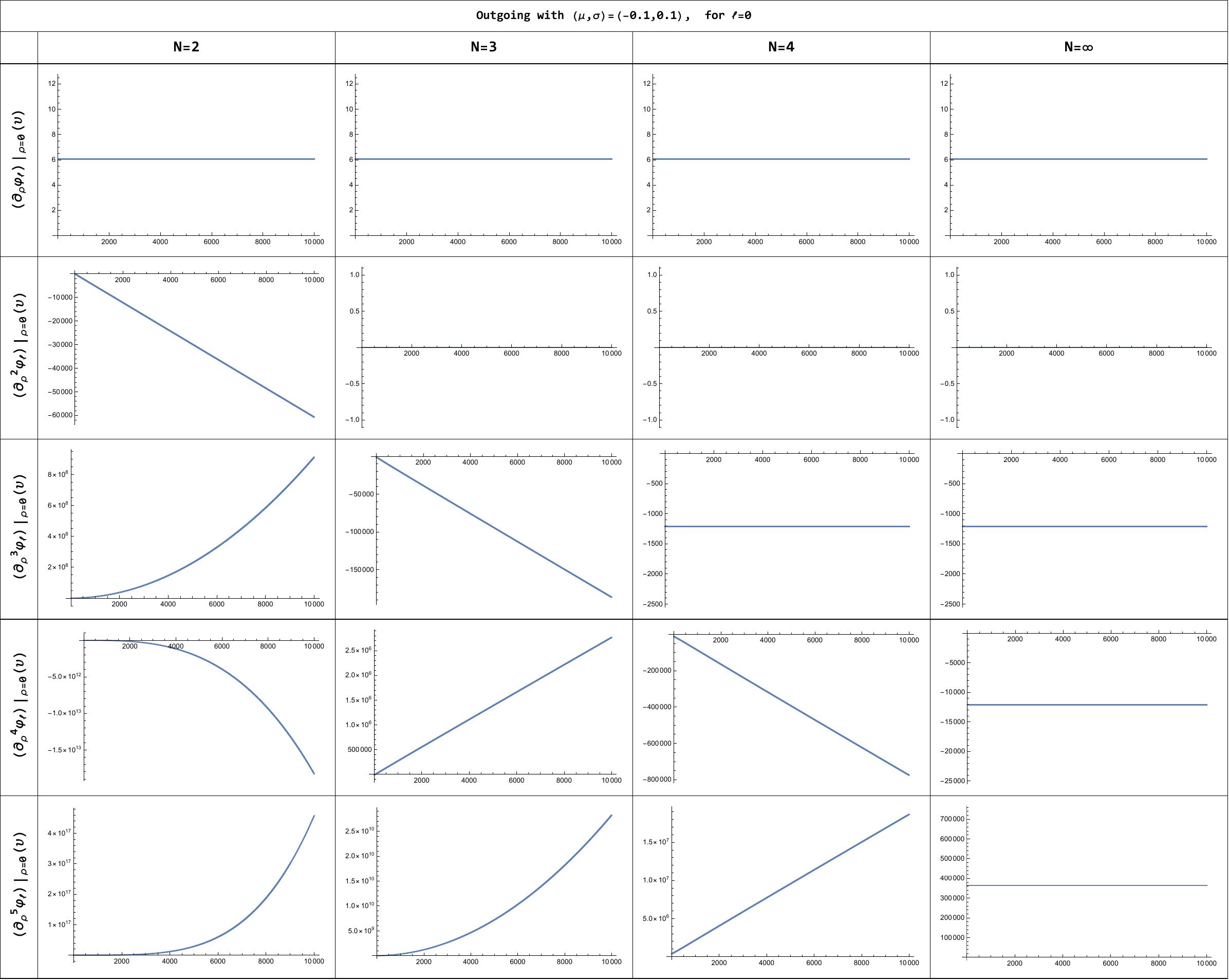}
    \caption{Time evolution of near-horizon transverse derivatives of radiative $s$-wave mode, $\left(\partial_{\rho}^{k}\varphi_{\ell=0}\right)\big|_{\rho=0}$, for ERN ($N=2$), multi-degenerate horizons with ranks of degeneracy $N=3$ and $N=4$, and an infinitely-degenerate horizon, starting from an initially outgoing wavepacket with $\left(\mu,\sigma\right)=\left(-0.1,0.1\right)$.}
    \label{fig:phiDrNdegL0}
\end{figure}

Comparing then the higher radial derivatives on the horizon for the s-wave mode, the pattern outlined in Eq.~\eqref{eq:FieldDerHorN} can be identified. As shown in Figure~\ref{fig:phiDrNdegL0}, the onset of linear blow-up is delayed by one derivative when the horizon degeneracy is increased by one, while the linear blow up persists for longer.

\begin{figure}[t]
    \centering
    \includegraphics[width=1.0\linewidth]{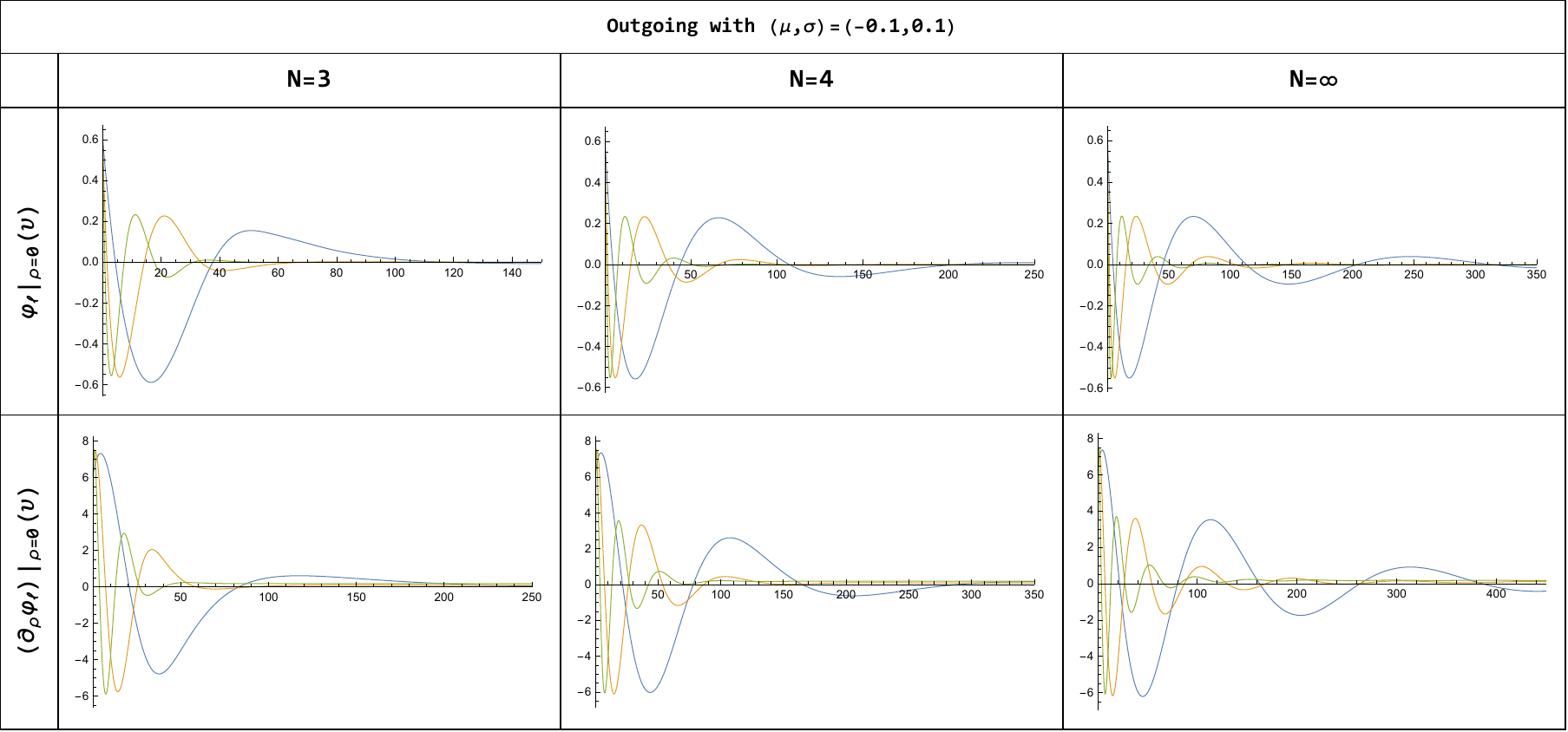}
    \caption{Time dependence of near-horizon projection of radiative modes and their first transverse derivative, $\varphi_{\ell}\big|_{\rho=0}$ and $\left(\partial_{\rho}\varphi_{\ell}\right)\big|_{\rho=0}$, starting from an outgoing wavepacket with $\left(\mu,\sigma\right)=\left(-0.1,0.1\right)$, for multi-degenerate horizons with ranks of degeneracy $N=3$ and $N=4$ and an infinitely-degenerate horizon. Each plot contains the evolution of the modes with orbital numbers $\ell=1$ (blue), $\ell=2$ (orange) and $\ell=3$ (green).}
    \label{fig:phiDrNdegL1}
\end{figure}

For spherical harmonic modes with $\ell \neq 0$, we argued that, for higher degeneracy horizons, the field and all its radial derivatives decay on the horizon at late advanced times. This can be checked for the first few terms in Figure~\ref{fig:phiDrNdegL1}.

To round up the results, we also include fields with initial ingoing profile (i.e. zero Aretakis constants), with the ingoing wavepacket characterized by parameters $\left(\mu^{\prime},\sigma^{\prime}\right)=\left(10.0,3.0\right)$. Figure~\ref{fig:phiDrNdegL0In} shows the behavior of the $s$-wave mode of the field and its radial derivatives on the horizon for different horizon degeneracies. A delay in the onset of instability is still observed with the increase in horizon degeneracy. The $\ell=1$ and $\ell=2$ modes are still seen to decay for higher degeneracy horizon as shown in Figure~\ref{fig:phiDrNdegL1In}.

\subsection{Infinitely-degenerate horizons}

The next configuration we will numerically analyze is an asymptotically flat spacetime that contains a single horizon with infinite rank of degeneracy. Specifically, we consider the geometry for which the areal radius is an affine parameter, $r\left(\rho\right)=\rho+r_{\text{h}}$, and with the metric discriminant function given by
\be
    f\left(\rho\right) = e^{-r_{\text{h}}^2/\rho^2} \,.
\ee

We use the same grid size and resolution as for the multiply-degenerate horizons, except for the simulation capturing the evolution of an $s$-wave outgoing wavepacket, for which the $\upsilon$-grid was selected to extend up to $\upsilon_{N_{\upsilon}}=1.0\cdot10^{5}$ and with resolution $N_{\upsilon}=5\cdot10^{5}$, so that the late time behavior is evident.

\begin{figure}[t]
    \centering
    \includegraphics[width=1.0\linewidth]{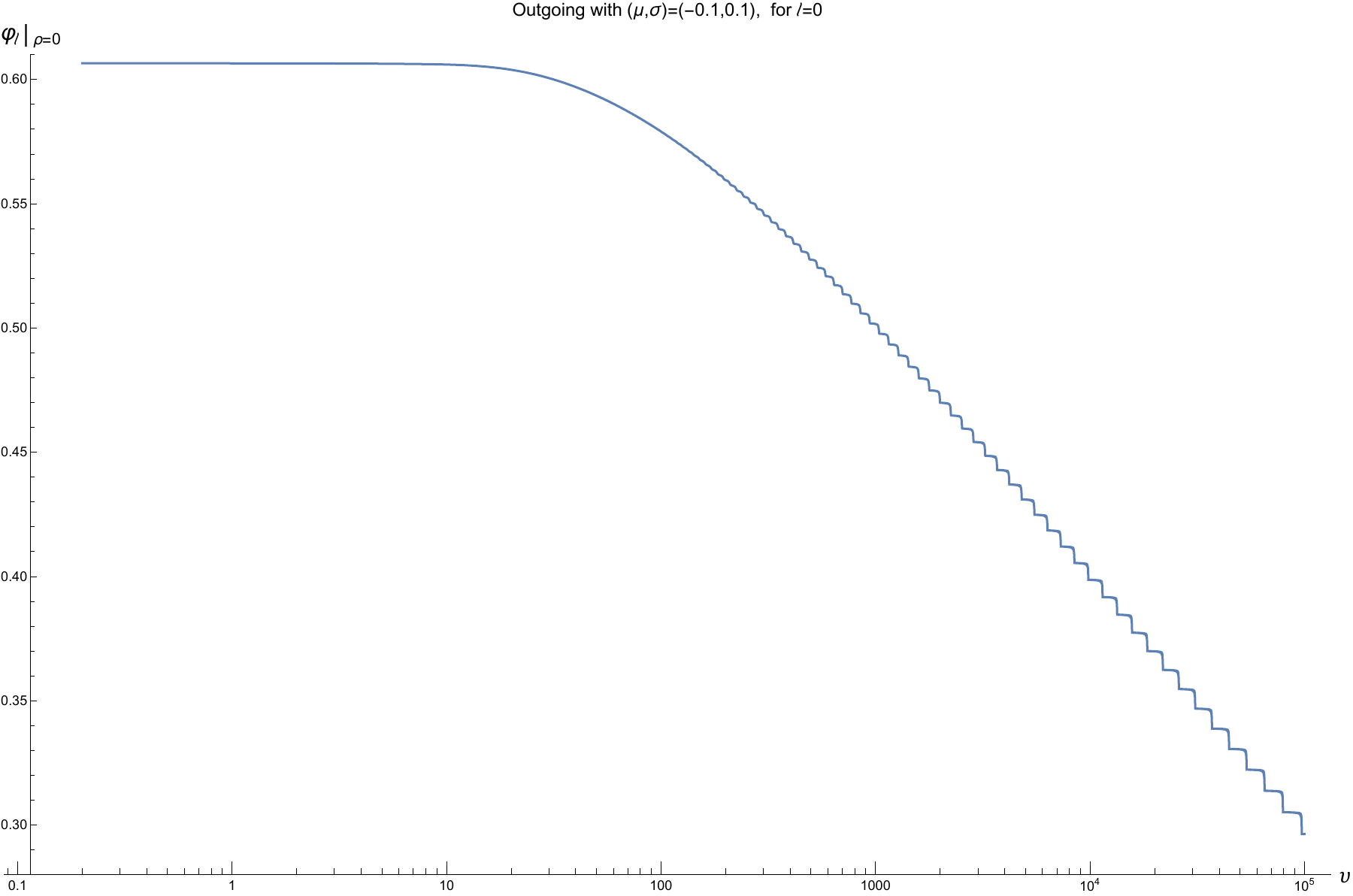}
    \caption{Time dependence of near-horizon radiative $s$-wave mode, $\varphi_{\ell=0}\big|_{\rho=0}$, for an initially outgoing wavepacket with $\left(\mu,\sigma\right)=\left(-0.1,0.1\right)$, in the background of an infinitely-degenerate horizon. The step-looking behavior is a numerical artifact due to the finite resolution along the $U$-direction.}
    \label{fig:phiInftydegL0}
\end{figure}

The evolution of an initially outgoing $s$-wave mode with $\left(\mu,\sigma\right)=\left(-0.1,0.1\right)$ is summarized in Figure~\ref{fig:phiDrNdegL0} and Figure~\ref{fig:phiInftydegL0}. Namely, Figure~\ref{fig:phiInftydegL0} confirms the decaying behavior of $\varphi_{\ell=0}\big|_{\rho=0}$ at late times. The last column of Figure~\ref{fig:phiDrNdegL0} is then in agreement with the analytic results of Section~\ref{sec:InfDegBH}, i.e. that the near-horizon transverse derivatives of the $s$-wave mode do not blow up at late times, approaching a non-zero constant that is a linear combinations of the $s$-wave Aretakis constants.

Next, the last column of Figure~\ref{fig:phiDrNdegL1} shows that both $\varphi_{\ell\ne0}\big|_{\rho=0}$ and $\left(\partial_{\rho}\varphi_{\ell\ne0}\right)\big|_{\rho=0}$ decay at late times, ensuring the absence of an Aretakis instability for an initially outgoing wavepacket with a non spherically symmetric profile ($\ell\ne0$).

\begin{figure}[t]
    \centering
    \includegraphics[width=1.0\linewidth]{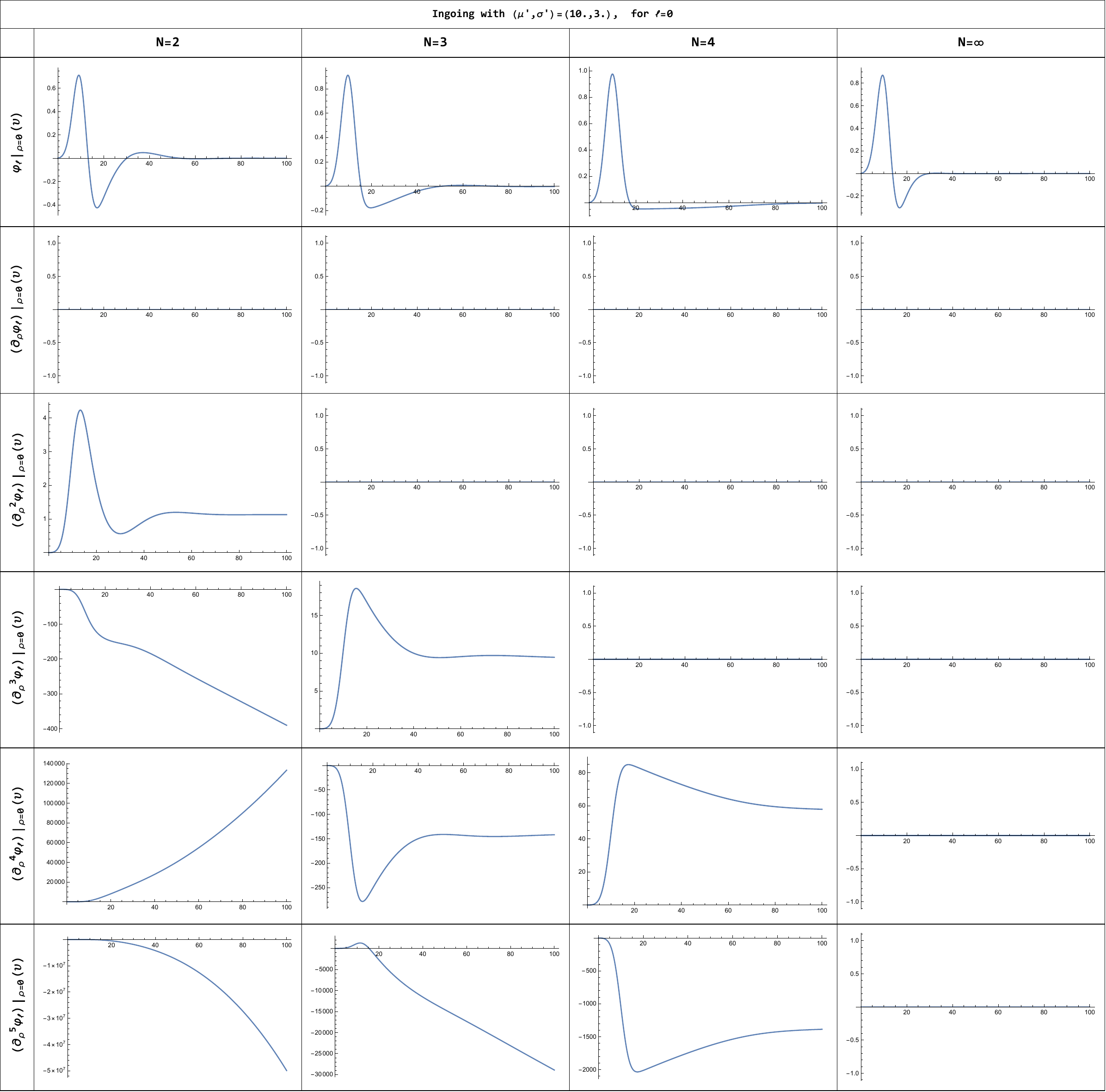}
    \caption{Time dependence of near-horizon field and its transverse derivatives for the $s$-wave mode, $\left(\partial_{\rho}^{k}\varphi_{\ell=0}\right)\big|_{\rho=0}$ for ERN ($N=2$), multi-degenerate horizons with ranks of degeneracy $N=3$ and $N=4$, and an infinitely-degenerate horizon, with the initial condition being an ingoing wavepacket with $\left(\mu^{\prime},\sigma^{\prime}\right)=\left(10.0,3.0\right)$.}
    \label{fig:phiDrNdegL0In}
\end{figure}

\begin{figure}
    \centering
    \includegraphics[width=1.0\linewidth]{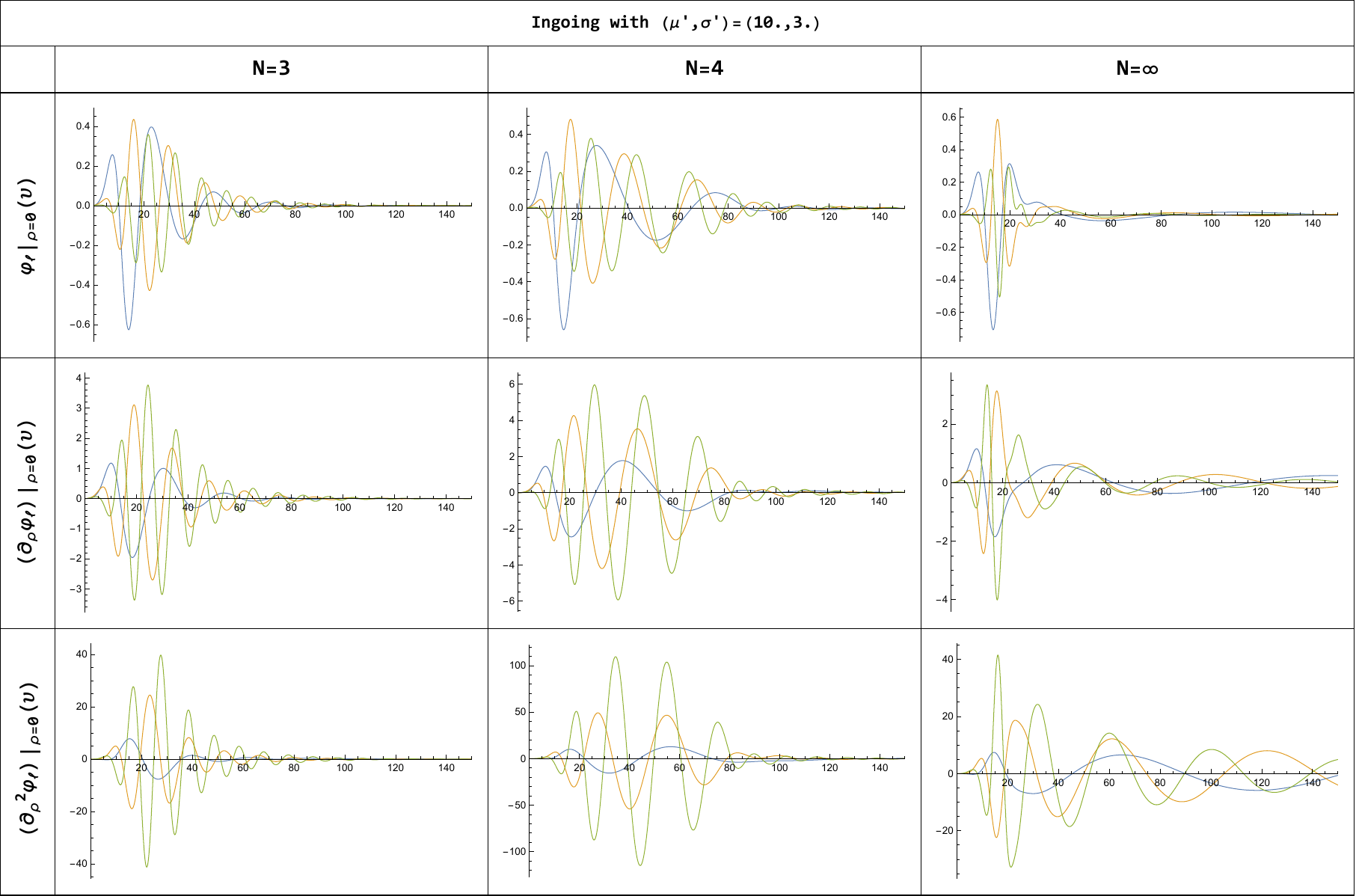}
    \caption{Time dependence of near-horizon field and its transverse derivatives, $\left(\partial_{\rho}^{k}\varphi_{\ell}\right)\big|_{\rho=0}$, for orbital numbers $\ell=1$ (blue), $\ell=2$ (orange) and $\ell=3$ (green), for multi-degenerate horizons with ranks of degeneracy $N=3$ and $N=4$, and an infinitely-degenerate horizon, with the initial condition being an ingoing wavepacket with $\left(\mu^{\prime},\sigma^{\prime}\right)=\left(10.0,3.0\right)$.}
    \label{fig:phiDrNdegL1In}
\end{figure}

As for the evolution of an initially ingoing wavepacket, Figure~\ref{fig:phiDrNdegL0In} and Figure~\ref{fig:phiDrNdegL1In} confirm the decay of both $\varphi_{\ell}\big|_{\rho=0}$ and all of the transverse derivatives $\left(\partial_{\rho}^{k\ge1}\varphi_{\ell}\right)\big|_{\rho=0}$, regardless of the value of $\ell$. The only special behavior is that, for the $s$-wave mode, all $\left(\partial_{\rho}^{k\ge1}\varphi_{\ell=0}\right)\big|_{\rho=0}$ vanish at all times due to the vanishing of the infinitely-many Aretakis constants associated with the infinitely-degenerate horizon.

\clearpage 
\section{Discussion and Outlook}
\label{sec:Discussion}

In this work, we revisited the role of the Aretakis instability within the broader problem of identifying dynamically viable endpoints of black hole evolution. Our motivation, as already emphasized in the Introduction, was that non-extremal inner horizons are generically plagued by mass inflation and semiclassical instabilities. As a result, extremal configurations may naturally emerge as preferred candidates for late-time endpoints, both because the Hawking temperature vanishes at zero surface gravity and because evolution beyond extremality is, in many settings, either forbidden or phenomenologically problematic. In this context, the Aretakis instability~\cite{Aretakis:2011ha,Aretakis:2011hc} becomes the main (classical) obstruction to the acceptance of degenerate horizons as genuinely stable end states. 

Our main result is that this obstruction is not universal in its strength, but depends crucially on the degree of degeneracy of the extremal black hole horizon. For multi-degenerate horizons of finite degeneracy rank $N$, the instability persists, but it is progressively screened as $N$ increases. More precisely, the late-time growth of transverse derivatives is postponed to increasingly higher orders, so that more derivatives approach constant values on the horizon before any blow-up sets in. In this sense, multi-degenerate horizons are indeed more stable than conventional doubly-degenerate horizons: the Aretakis mechanism is still present, but in a milder form. This conclusion, initially derived analytically, is also supported by the numerical evolutions we presented in Section~\ref{sec:NumericalResults}. 

This softening admits a simple physical interpretation. A central lesson from the extremal Reissner--Nordstr\"om case, reviewed in Section~\ref{sec:AretakisReview}, is that the Aretakis instability is controlled entirely by the near-horizon geometry. It is therefore a local phenomenon, largely insensitive to the far-region structure of the spacetime. The same lesson applies here. What matters for the instability is not whether the geometry globally resembles ERN or some other extremal solution, but rather how the metric behaves arbitrarily close to the horizon. This observation opens an interesting possibility: we may be able to construct geometries that are essentially indistinguishable from ordinary (vacuum solutions of GR) extremal black holes throughout most of the exterior region, while differing only within a narrow layer near the horizon, where higher-order degeneracy radically alters the Aretakis instability, by softening or even removing it altogether.

Motivated by these results, we proposed a class of geometries endowed with an infinitely-degenerate horizon. The explicit example studied in Section~\ref{sec:InfDegBH} should be viewed only as one representative of a broader class, rather than a unique construction. What is crucial is the existence of infinitely many horizon conservation laws in the $s$-wave sector, which prevent any transverse derivative from developing the polynomial late-time growth characteristic of the Aretakis instability. Together with our numerical results, this strongly suggests that infinitely-degenerate horizons may provide classically stable configurations, at least against the Aretakis-type perturbations studied here. In this sense, they offer a concrete realization of the ``black-hole graveyard'' idea~\cite{DiFilippo:2024spj}: geometries that evade mass inflation, suppress Hawking evaporation, and at the same time avoid the standard classical instability of extremal horizons. 

At the same time, our results should be interpreted with appropriate caution. First, the present analysis is restricted to linear probe perturbations, and mostly to static, spherically symmetric backgrounds. Although the near-horizon scaling arguments suggest a degree of universality, extending the discussion to fully nonlinear dynamics remains essential. In this respect, it would be particularly interesting to clarify the relation between our picture and recent analytic work on the late-time near-horizon dynamics of extremal black holes~\cite{Porfyriadis:2025pov}. Second, while our infinitely-degenerate examples appear free from an Aretakis blow-up, a full assessment of their viability also requires understanding whether other kinds of horizon pathologies or long-lived excitations may arise in their place. In particular, it would be worthwhile to investigate the spectrum of quasinormal modes, as well as signatures in photon-ring structure, ringdown, absorption properties, and the response to external perturbations. Even if the near-horizon modification is confined to a parametrically small region, it may still leave an imprint through late-time tails or through observables especially sensitive to the throat geometry. Establishing whether such signatures are measurable remains an open question, but one that appears to be directly tied to the properties of locality emphasized throughout this work.

Several extensions of the present work appear especially promising. The most immediate is the generalization to rotating black holes, where the interplay among extremality, superradiance, inner-horizon structure and Aretakis-type phenomena is expected to be considerably richer. It would also be important to understand whether multi-degenerate or infinitely-degenerate rotating horizons can be constructed in a controlled way, and whether the corresponding near-horizon geometries similarly screen or eliminate the instability. Finally, recent developments on quantum fluctuations in near-extremal black holes suggest that the classical picture of horizon instabilities may need to be revisited in a broader framework~\cite{Iliesiu:2020qvm,Emparan:2025qqf}. Clarifying whether the infinitely-degenerate horizons  retain their stability in these settings may help determine whether they are merely mathematical curiosities or genuine, physically relevant, endpoint configurations. We leave these questions to future investigations.

\paragraph{Acknowledgments}
S.A., P.C. and L.D. are supported by the European Research Council (ERC) Project 101076737 -- CeleBH. Views and opinions expressed are however those of the authors only and do not necessarily reflect those of the European Union or the European Research Council. Neither the European Union nor the granting authority can be held responsible for them. S.A., P.C. and L.D. are also partially supported by the INFN Iniziativa Specifica ST\&FI. S.L and G.N. are also partially supported by the INFN Iniziativa Specifica QUAGRAP.

\appendix

\section{Aretakis instability for multi-degenerate horizons}
\label{app:AretakisMultiDegenerate}

In this appendix, we derive the form of the Aretakis instability that emerges in the $s$-wave sector of massless scalar probe fields in the background of a static and spherically symmetric spacetimes that is equipped with a multi-degenerate horizon. In horizon-centered advanced null Gaussian coordinates $\left(\upsilon,\rho,x^{A}\right)$, the metric near a horizon of degeneracy rank $N$ is given by Eq.~\eqref{eq:MetricHorN}. The corresponding equation of motion of interest, $\Box\Phi=0$, reduces to Eq.~\eqref{eq:boxNdeg} after expanding the scalar field into spherical harmonic modes $\Phi_{\ell m}$. For the $s$-wave ($\ell=0=m$) mode, in particular, this simplifies to
\be\label{eq:boxNdegS}
	\partial_{\upsilon}\left(r\partial_{\rho}\left(r\Phi_{00}\right)\right)=-\frac{1}{2}\partial_{\rho}\left(\rho^{N}r^2b\,\partial_{\rho}\Phi_{00}\right) \,.
\ee
As already remarked in Section~\ref{sec:AretakisHorN}, taking the near-horizon limit, $\rho\rightarrow0$, reveals the existence of the following $N-1$ Aretakis constants,
\be
	\begin{gathered}
		A_{00}^{\left(n\right)} = \frac{r_{\text{h}}^{n-1}}{\left(n+1\right)!}\lim_{\rho\rightarrow0}\partial_{\rho}^{n}\left[r\partial_{\rho}\left(r\Phi_{00}\right)\right] \\
		\Rightarrow \quad \partial_{\upsilon}A_{00}^{\left(n\right)} = 0 \,,\quad 0\le n\le N-2 \,.
	\end{gathered}
\ee

We will now examine the implications of the existence of these constants on the stability or instability of the $s$-wave mode. For the scope of this, let us denote the dimensionless near-horizon transverse derivatives of the $s$-wave mode as
\be
	\phi_{n}(\upsilon)\coloneqq \frac{r_{\text{h}}^{n}}{n!}\left(\partial_{\rho}^{n}\Phi_{00}\right)\big|_{\rho=0} \,,
\ee
and let us also introduce the dimensionless coefficients
\be
	R_{n}\coloneqq \frac{r_{\text{h}}^{n-2}}{n!}\left(\partial_{\rho}^{n}r^2\right)\big|_{\rho=0} \,.
\ee
Then, the $s$-wave Aretakis constants are rewritten as
\be\label{eq:A00kConstantsValues}
	A_{00}^{\left(n\right)} = \phi_{n+1} +\sum_{m=0}^{n}\frac{n+m+1}{2\left(n+1\right)}R_{n-m+1}\phi_{m} \,,
\ee
which can be viewed as a recurrence relation that allows to express all $\phi_{k}$, with $1\le k\le N-1$, in terms of $\phi_0=\Phi_{00}|_{\rho=0}$ and linear combinations of $A_{00}^{\left(n\le k-1\right)}$. Therefore, assuming that $\phi_{0}$ decays at late advanced times, we see that the first $N-1$ transverse derivatives $\phi_{1\le k\le N-1}$ do \textit{not} decay for non-zero Aretakis constants.

For instance, for any $N\ge 3$, we can write the first two transverse derivatives on the horizon as
\be\ba
	\phi_1 &= A_{00}^{\left(0\right)} -\frac{1}{2}R_1\phi_0 \,, \\
	\phi_2 &= A_{00}^{\left(1\right)} -\frac{3}{4}R_1A_{00}^{(0)} -\frac{1}{2}\left(R_2-\frac{3}{4}R_1^2\right)\phi_0 \,.
    \\
\ea\ee
whence it follows, assuming that $\phi_0\to 0$ at late times, that
\be\ba\label{eq:NdegPhinLateTime}
	\phi_1 &\xrightarrow{\upsilon\rightarrow\infty} A_{00}^{\left(0\right)} \,, \\
	\phi_2 &\xrightarrow{\upsilon\rightarrow\infty} A_{00}^{\left(1\right)} -\frac{3}{4}R_1A_{00}^{(0)} \,.
\ea\ee

Next, acting $k\ge N-1$ times with $\partial_{\rho}$ onto Eq.~\eqref{eq:boxNdegS} and evaluating on the horizon implies that
\be\label{eq:DkNdegWaveEq}
	\partial_{\upsilon}\left(\partial_{\rho}^{k}\left[r\partial_{\rho}(r\Phi_{00})\right]\right)\big|_{\rho=0} = -\frac{1}{2}\sum_{n=1}^{k-N+2}\frac{\left(k+1\right)!}{\left(n-1\right)!}\frac{c_{k-n-N+2}}{r_{\text{h}}^{k-n}}\left(\partial_{\rho}^{n}\Phi_{00}\right)\big|_{\rho=0} \,,
\ee
where we introduced the dimensionless coefficients of the near-horizon expansion of the function $r^2b$,
\be
	c_{n} \coloneqq \frac{r_{\text{h}}^{n+N-2}}{n!}\left(\partial_{\rho}^{n}(r^2b)\right)\big|_{\rho=0} \,.
\ee
In terms of the previously introduced field coefficients, we can rearrange Eq.~\eqref{eq:DkNdegWaveEq} into
\be\label{eq:DkNdegWaveEq2}
	\partial_{\upsilon}\phi_{k+1} = -\frac{1}{2}\bigg\{\sum_{n=1}^{k-N+2}nc_{k-n-N+2}\frac{\phi_{n}}{r_{\text{h}}}  +\sum_{n=0}^{k}\frac{k+n+1}{k+1}R_{k-n+1}\partial_{\upsilon}\phi_{n}\bigg\} \,.
\ee

Now, setting $k=N-1$, we get
\be
	\partial_{\upsilon}\phi_{N} = -\frac{1}{2}\bigg\{c_{0}\frac{\phi_1}{r_{\text{h}}} +\sum_{n=0}^{N-1}\frac{n+N}{N}R_{N-n}\partial_{\upsilon}\phi_{n}\bigg\} \,,
\ee
whose RHS is dominated by the first term at late times, as any of the involved $\phi_{n}$ is either assumed to vanish ($n=0$) or approaches a constant ($1\le n\le N-1$). As the example in Eq.~\eqref{eq:NdegPhinLateTime} shows, the only remaining contribution will be proportional to $A_{00}^{\left(0\right)}$. As a result, $\phi_{N}$ must blow up linearly at large $\upsilon$,
\be\ba\label{eq:NdegPhinLateTimeN}
	\partial_{\upsilon}\phi_{N} \xrightarrow{\upsilon\rightarrow\infty} -\frac{c_0}{2r_{\text{h}}}A_{00}^{(0)} \quad\Rightarrow\quad \phi_{N} \xrightarrow{\upsilon\rightarrow\infty} -A_{00}^{\left(0\right)}\frac{c_0\upsilon}{2r_{\text{h}}} \,.
\ea\ee

Next, setting $k=N$ in Eq.~\eqref{eq:DkNdegWaveEq2} and taking the large $\upsilon$ limit, we realize that the RHS is dominated at most by three terms
\be
	\partial_{\upsilon}\phi_{N+1} \xrightarrow{\upsilon\rightarrow\infty} -\frac{1}{r_{\text{h}}}\bigg\{ c_0\phi_2 +\frac{1}{2}c_1\phi_1 +\frac{2N+1}{2\left(N+1\right)}R_1r_{\text{h}}\partial_{\upsilon}\phi_{N} \bigg\} \,.
\ee
For $N=2$, only the first term dominates and blows up linearly in $\upsilon$, implying that $\phi_3$ must blow up as $\upsilon^2$ at large $\upsilon$. This is the usual Aretakis instability we reviewed in Section~\ref{sec:AretakisReview} (see Eq.~\eqref{eq:AretakisInstabilityConventional}). For $N\ge3$, on the other hand, all three terms in the RHS become constants at late times. The specific value is related to the Aretakis constants $A_{00}^{\left(0\right)}$ and $A_{00}^{\left(1\right)}$, as per Eq.~\eqref{eq:A00kConstantsValues}. Using Eq.~\eqref{eq:NdegPhinLateTime} and Eq.~\eqref{eq:NdegPhinLateTimeN}, we can work out that
\be
	\partial_{\upsilon}\phi_{N+1} \xrightarrow{\upsilon\rightarrow\infty} -\frac{c_0}{2r_{\text{h}}}\mathcal{A}_{N,1}^{\left(1\right)} \quad\Rightarrow\quad \phi_{N+1} \xrightarrow{\upsilon\rightarrow\infty} -\mathcal{A}_{N,1}^{\left(1\right)}\frac{c_0\upsilon}{2r_{\text{h}}} \quad\text{for $N\ge3$} \,,
\ee
with
\be
	\mathcal{A}_{N,1}^{\left(1\right)} \coloneqq 2A_{00}^{\left(1\right)} +\left[\frac{c_1}{c_0}-\frac{5N+4}{2\left(N+1\right)}R_1\right]A_{00}^{(0)} \,.
\ee
For the next transverse derivative, one studies the late time behavior of Eq.~\eqref{eq:DkNdegWaveEq2} after setting $k=N+1$ to analogously find that $\phi_{N+2}$ behaves as
\be
	\phi_{N+2} \xrightarrow{\upsilon\rightarrow\infty}
	\begin{cases}
		A_{00}^{\left(0\right)}\left(-\frac{c_0\upsilon}{2r_{\text{h}}}\right)^3 & \text{if $N=2$} \,; \\
		\frac{3}{2}A_{00}^{\left(0\right)}\left(-\frac{c_0\upsilon}{2r_{\text{h}}}\right)^2 & \text{if $N=3$} \,; \\
		\mathcal{A}_{N,1}^{\left(2\right)}\left(-\frac{c_0\upsilon}{2r_{\text{h}}}\right) & \text{if $N\ge4$} \,.
	\end{cases}
\ee
with a constant coefficient that now depends on $A_{00}^{\left(2\right)}$ as well
\be\ba
	{}&\mathcal{A}_{N,1}^{\left(2\right)} \coloneqq 3A_{00}^{\left(2\right)} +2\left[\frac{c_1}{c_0}-\frac{17N+28}{12\left(N+2\right)}R_1\right]A_{00}^{\left(1\right)} \\
	&\quad +\left[\frac{c_2}{c_0}-\frac{5N+9}{2\left(N+2\right)}R_1\frac{c_1}{c_0} -\frac{3N+5}{N+2}R_2+\frac{35N^2+91N+54}{8\left(N+1\right)\left(N+2\right)}R_1^2\right]A_{00}^{(0)} \,.
\ea\ee
Moving on, setting $k=N+2$ in Eq.~\eqref{eq:DkNdegWaveEq2} and following the same procedure, we find the following qualitative late time behavior
\be
	\phi_{N+3} \xrightarrow{\upsilon\rightarrow\infty}
	\begin{cases}
		A_{00}^{(0)}\left(-\frac{c_0\upsilon}{2r_{\text{h}}}\right)^4 & \text{if $N=2$} \,; \\
		\mathcal{A}_{3,2}^{\left(1\right)}\left(-\frac{c_0\upsilon}{2r_{\text{h}}}\right)^2 & \text{if $N=3$} \,; \\
		2A_{00}^{(0)}\left(-\frac{c_0\upsilon}{2r_{\text{h}}}\right)^2 & \text{if $N=4$} \,; \\
		\mathcal{A}_{N,1}^{(3)}\left(-\frac{c_0\upsilon}{2r_{\text{h}}}\right) & \text{if $N\ge5$} \,.
	\end{cases}
\ee
with
\be
	\mathcal{A}_{3,2}^{\left(1\right)} = 4A_{00}^{\left(1\right)} +\frac{10}{3}\left[\frac{c_1}{c_0}-\frac{107}{40}R_1\right]A_{00}^{\left(0\right)} \,,
\ee
and a rather lengthy expression of $\mathcal{A}_{N,1}^{\left(3\right)}$ in terms of $A_{00}^{\left(n\right)}$, $c_{n}/c_0$ and $R_{n}$, with $n\le3$, that we omit. Repeating this procedure for larger values of $k$ reveals a clear pattern for the higher transverse derivatives of the $s$-wave perturbation modes on a $N$-degenerate horizon, which can be confirmed by induction:
\be
	\phi_{k} \xrightarrow{\upsilon\rightarrow\infty}
	\begin{cases}
		0 & \text{for $k=0$ (assumption)} \,; \\
		\mathcal{A}_{N,0}^{\left(k-1\right)} & \text{for $1\le k\le N-1$} \,; \\
		\mathcal{A}_{N,1}^{\left(k-N\right)}\left(-\frac{c_0\upsilon}{2r_{\text{h}}}\right) & \text{for $N\le k\le 2N-2$} \,; \\
		\mathcal{A}_{N,2}^{\left(k-2N+1\right)}\left(-\frac{c_0\upsilon}{2r_{\text{h}}}\right)^2 & \text{for $2N-1\le k\le 3N-3$} \,; \\
		\mathcal{A}_{N,3}^{\left(k-3N+2\right)}\left(-\frac{c_0\upsilon}{2r_{\text{h}}}\right)^3 & \text{for $3N-2\le k\le 4N-4$} \,; \\
		\vdots & \vdots \\
		\mathcal{A}_{N,p}^{\left(k-p\left(N-1\right)-1\right)}\left(-\frac{c_0\upsilon}{2r_{\text{h}}}\right)^{p} & \text{for $p\left(N-1\right)+1\le k\le \left(p+1\right)\left(N-1\right)$} \,; \\
		\vdots & \vdots
	\end{cases}
\ee
In the above expressions, $\mathcal{A}_{N,p}^{\left(k\right)}$ are linear combinations of $A_{00}^{\left(n\le k\right)}$, and also non-linear combinations of $c_{n\le k}$ and $R_{n\le k}$, whose explicit expression can be found case-by-case by solving the recursion relations of Eq.~\eqref{eq:DkNdegWaveEq2} in the relevant range of indices. For instance, $\mathcal{A}_{N,0}^{\left(k\right)} = \sum_{n=1}^{k}\left(\mathcal{R}^{-1}\right)_{k,n}A_{00}^{\left(n-1\right)}$ with $\mathcal{R}$ the triangular matrix with entries
\be
	\mathcal{R}_{n,m} = \frac{n+m}{2n}R_{n-m}\delta_{1\le m\le n} \,.
\ee
More compactly, if $\phi_0$ decays at late times and $A_{00}^{\left(n\right)}\ne0$, then
\be
	\phi_{k\ge1} \propto \upsilon^{\left\lceil\frac{k}{N-1}\right\rceil-1} \quad\text{as $\upsilon\rightarrow\infty$} \,.
\ee
In fact, this conclusion on the late-time behavior of $\phi_{k\ge1}$ remains true even if $\phi_0$ approaches a constant at late times, as can be checked explicitly by repeating the aforementioned procedure, the only complication being a more involved proportionality constant $\mathcal{A}_{N,p}^{\left(k\right)}$ that depends linearly on $\lim_{\upsilon\rightarrow\infty}\phi_0$.


\section{Photon sphere stability}
\label{app:PhotonSphereStability}

In this appendix, we study null geodesics of a generic static and spherically symmetric spacetime, which can always be brought to the form
\be
	ds^2 = -f\left(\rho\right)\dd\upsilon^2 +2\dd\upsilon\dd\rho +r^2\left(\rho\right)\left(\dd\theta^2+\sin^2\theta\dd\phi^2\right)
\ee
in advanced null Gaussian coordinates. There are in general four Killing vectors associated with such geometries, one for the time translation symmetry and three for the spherical symmetry, given explicitly by
\be\ba
	\xi_{\left(0\right)} &= \partial_{\upsilon} \,,\quad \xi_{\left(3\right)} = \partial_{\phi} \,, \\
	\xi_{\left(1\right)} &= -\cos\phi\,\partial_{\theta}+\cot\theta\sin\phi\,\partial_{\phi} \,, \\
	\xi_{\left(2\right)} &= \sin\phi\,\partial_{\theta}+\cot\theta\cos\phi\,\partial_{\phi} \,,
\ea\ee
and spanning a $\mathfrak{u}\left(1\right)\times\mathfrak{so}\left(3\right)$ algebra.

We will be dealing with geometries that contain a horizon placed at the origin of the radial null Gaussian coordinate $\rho$, i.e.
\be
	f\left(\rho=0\right) = 0 \,.
\ee
In particular, if this is the event horizon of a black hole, generated by the Killing vector $\partial_{\upsilon}$, the surface gravity $\kappa$ is obtained according to
\be
	\kappa = \frac{1}{2}f^{\prime}\left(0\right) \Rightarrow f\left(\rho\right) = 2\kappa\rho +\mathcal{O}\left(\rho^2\right) \,.
\ee


\subsection{Electrically neutral geodesics}

Electrically neutral geodesics follow from the effective Lagrangian
\be
	\mathcal{L} = \frac{1}{2}g_{\mu\nu}\dot{x}^{\mu}\dot{x}^{\nu} = -\frac{1}{2}f\left(\rho\right)\dot{\upsilon}^2 +\dot{\upsilon}\dot{\rho} +\frac{1}{2}r^2\left(\rho\right)\left(\dot{\theta}^2+\sin^2\theta\dot{\phi}^2\right) \,,
\ee
where $\dot{x}^{\mu} \coloneqq \frac{dx^{\mu}}{ds}$, $s$ being an affine parameter for the geodesic. The associated conjugate momentum is given by
\be
	p_{\mu} = \frac{\partial\mathcal{L}}{\partial\dot{x}^{\mu}} = g_{\mu\nu}\dot{x}^{\nu} \,.
\ee
Each of the four Killing vectors gives rise one integrals of motion: the orbital energy $E\coloneqq-\xi_{\left(0\right)}^{\mu}p_{\mu}$ and the three components of the angular momenta $L_{i}\coloneqq\xi_{\left(i\right)}^{\mu}p_{\mu}$, $i=1,2,3$,
\be\ba
	E &= f\left(\rho\right)\dot{\upsilon}-\dot{\rho} \,,\quad L_3=r^2\sin^2\theta\,\dot{\phi} \\
	L_{1} &= r^2\left(-\cos\phi\,\dot{\theta}+\sin\theta\cos\theta\sin\phi\,\dot{\phi}\right) \,, \\
	L_{2} &= r^2\left(\sin\phi\,\dot{\theta}+\sin\theta\cos\theta\cos\phi\,\dot{\phi}\right) \,. \\
\ea\ee
Furthermore, the effective Hamiltonian, $\mathcal{H}=p_{\mu}\dot{x}^{\mu}-\mathcal{L}=\frac{1}{2}p_{\mu}p^{\mu}$, is identified with the (unit) inertial mass-squared, $\mathcal{H}=-\frac{\sigma}{2}$, with $\sigma=+1,0,-1$ for timelike, null and spacelike geodesics respectively. Putting everything together, the geodesic motion reduces to the radial problem
\be
	\dot{\rho}^2 +\mathcal{V}_{\text{eff}}\left(\rho\right) = 0 \,,
\ee
with the effective potential given by
\be
	\mathcal{V}_{\text{eff}}\left(\rho\right) = -E^2 +f\left(\rho\right)\left(\frac{L^2}{r^2\left(\rho\right)} +\sigma\right) \,,
\ee
where $L^2$ above is the total angular momentum
\be
	L^2 \coloneqq L_1^2 +L_2^2 +L_3^2 \,,
\ee
the important point being that it is a constant of motion.


\subsection{Photon spheres}

Let us now focus to the study of photon spheres, i.e. null geodesics ($\sigma=0$) of constant radius $\rho=\rho_{\text{ph}}$, for which
\be
	\mathcal{V}_{\text{eff}}\left(\rho_{\text{ph}}\right) = 0 = \mathcal{V}_{\text{eff}}^{\prime}\left(\rho_{\text{ph}}\right) \,.
\ee
If the horizon at $\rho=0$ is non-degenerate ($\kappa\ne0$), it is not hard to see that, near extremality ($\kappa r_{\text{h}}\ll1$, with $r_{\text{h}}\coloneqq r\left(\rho=0\right)$) there is a photon sphere located at
\be
	\rho_{\text{ph}} = \frac{2\kappa}{f^{\prime\prime}\left(0\right)} +\mathcal{O}\left(\kappa^2\right) \,.
\ee

More interestingly, when the horizon is degenerate, $\kappa=0$, then we reach the rather universal result that there always exists a photon sphere located on the horizon,
\be
	\text{If $\kappa=0$} \quad\Rightarrow\quad \rho_{\text{ph}} = 0 = \rho_{\text{horizon}} \,.
\ee
It is straightforward to generalize this statement to degenerate horizons of higher rank of degeneracy. Namely, if the horizon has a degeneracy of rank $N$, that is, if
\be
	f\left(\rho\right) = \rho^{N}b\left(\rho\right) \,,\quad N\ge2 \,,
\ee
with $b\left(0\right)\ne0$, then
\be
	\mathcal{V}_{\text{eff}}^{\prime}\left(\rho\right) = \rho^{N-1}\frac{Nb\left(0\right)}{r_{\text{h}}^2} + \mathcal{O}\left(\rho^{N}\right) \,,
\ee
which confirms that there is a photon sphere located on the degenerate horizon for any rank of degeneracy $N\ge2$. From the additional defining condition $\mathcal{V}_{\text{eff}}\left(\rho_{\text{ph}}\right) = 0$, one furthermore notices that such null geodesics have zero orbital energy,
\be
	E\big|_{\rho_{\text{ph}}=0}=0 \,.
\ee

Last, let us examine the stability properties of this photon sphere. From the behavior of the effective potential near the photon sphere, we see that the photon sphere on the degenerate horizon is stable for even $N$ and meta-stable for odd $N$. This involves the positivity observation that
\be
	\mathcal{V}_{\text{eff}}^{\left(N\right)}\left(\rho_{\text{ph}}=0\right) = \frac{N!b\left(0\right)}{r_{\text{h}}^2} >0 \,,
\ee
by virtue of causality, i.e. that $b\left(0\right)>0$, such that $g_{\upsilon\upsilon}<0$ in the exterior ($\rho>0$).

The above stability argument for even ranks of degeneracy relies on the assumption that the horizon of interest is an outer horizon. If, instead, the horizon at $\rho=0$ is inside a series of $k$ horizons, with the $i$'th horizon having degeneracy rank $N_{i}$, then one instead has
\be
	\text{sign}\left\{g_{\upsilon\upsilon}\right\} = \prod_{i=1}^{k}\left(-1\right)^{N_{i}} \quad \text{as $\rho\rightarrow0^{+}$} \,.
\ee
As such, the photon sphere on the $N$-tuply degenerate horizon at $\rho=0$ is stable if $N$ is even and $\sum_{i=1}^{k}N_{i}$ is also even, and it is unstable if $N$ is even and $\sum_{i=1}^{k}N_{i}$ is odd, while it is meta-stable if $N$ is odd.


\section{Relation between $\rho$ and $\left(\upsilon,U\right)$}
\label{app:rhoUv}

In this appendix, we study the relation between the radial, horizon-centered, null Gaussian coordinate $\rho$ and the double null coordinates $\left(\upsilon,U\right)$ that are regular on the horizon, introduced in Section~\ref{sec:NumericalResults}. The aim is to solve the algebraic equation
\be\label{eq:rhoUvDef}
    r_{\ast}\left(\rho\left(\upsilon,U\right)\right) = \frac{\upsilon-\upsilon_0}{2}+r_{\ast}\left(-U\right) \,,
\ee
with $r_{\ast}$ the tortoise mapping, in an appropriate expansion near $U=0$. The tortoise mapping for geometries of the form $ds^2 = -f\left(\rho\right)\dd\upsilon^2+2\dd\upsilon\dd\rho +r^2\left(\rho\right)\dd\Omega_2^2$ is given by
\be
    r_{\ast} : \rho \mapsto \int^{\rho}\frac{d\rho^{\prime}}{f\left(\rho^{\prime}\right)} \,,
\ee
up to an arbitrary integration constant.

The first observation is that, near $U=0$, the term $\frac{\upsilon-\upsilon_0}{2}$ in Eq.~\eqref{eq:rhoUvDef} is suppressed at finite advanced time instants, giving the universal leading order behavior
\be
    \rho\left(\upsilon,U\right) = -U +r_{\text{h}}\varepsilon\left(\upsilon,U\right) \,,\quad \left|\varepsilon\left(\upsilon,U\right)\right|\ll \left|\frac{U}{r_{\text{h}}}\right| \,,
\ee
where $r_{\text{h}}\coloneqq r\left(\rho=0\right)$ is the horizon areal radius, introduced to work with manifestly dimensionless quantities.

Using this, we expand $r_{\ast}\left(\rho\left(\upsilon,U\right)\right)$ in powers of $\varepsilon\left(\upsilon,U\right)$
\be
    r_{\ast}\left(\rho\left(\upsilon,U\right)\right) = r_{\ast}\left(-U\right) + r_{\text{h}}\sum_{n=1}^{\infty}C_{n}\left(-U\right)\varepsilon^{n}\left(\upsilon,U\right) \,,
\ee
with
\be\label{eq:CnDef}
    C_{n}\left(\rho\right)\coloneqq \frac{r_{\text{h}}^{n-1}}{n!}\frac{d^{n}}{d\rho^{n}}r_{\ast}\left(\rho\right) = \frac{r_{\text{h}}^{n-1}}{n!}\frac{d^{n-1}}{d\rho^{n-1}}\frac{1}{f\left(\rho\right)} \,,
\ee
where we used the geometric fact that $r_{\ast}^{\prime}\left(\rho\right)=\frac{1}{f\left(\rho\right)}$. Then, Eq.~\eqref{eq:rhoUvDef} is rewritten to
\be
    \sum_{n=1}^{\infty}C_{n}\left(-U\right)\varepsilon^{n}\left(\upsilon,U\right) = \frac{\upsilon-\upsilon_0}{2r_{\text{h}}} \,.
\ee
Using the Lagrange inversion theorem, this admits the following asymptotic series expansion
\be\label{eq:rhoUvModes}
    \varepsilon\left(\upsilon,U\right) \sim \sum_{n=1}^{\infty}\left(\frac{\upsilon-\upsilon_0}{2r_{\text{h}}C_1\left(-U\right)}\right)^{n}\mathcal{P}_{n}\left(F_1\left(-U\right),F_2\left(-U\right),\dots,F_{n-1}\left(-U\right)\right) \,,
\ee
where $\mathcal{P}_{n}\left(x_1,x_2,\dots,x_{n-1}\right)$ is a polynomial constructed from the following sum of incomplete exponential Bell polynomials\footnote{The incomplete exponential Bell polynomial $B_{n,k}\left(x_1,x_2,\dots,x_{n-k+1}\right)$ is given by
\begin{equation*}
    B_{n,k}\left(x_1,x_2,\dots,x_{n-k+1}\right) = n!\sum \prod_{m=1}^{n-k+1}\frac{x_{m}^{j_{m}}}{\left(m!\right)^{j_{m}}j_{m}!} \,,
\end{equation*}
where the multi-sum is taken over all possible integer values of $j_{m}\ge0$ satisfying
\begin{equation*}
    \sum_{m=1}^{n-k+1}j_{m} = k \quad\text{and}\quad \sum_{m=1}^{n-k+1}mj_{m} = n \,.
\end{equation*}
}
\be\label{eq:PnBnk}
    \mathcal{P}_{n}\left(x_1,x_2,\dots,x_{n-1}\right) = \sum_{k=1}^{n-1}\left(-1\right)^{k}\frac{\left(n+k-1\right)!}{n!\left(n-1\right)!}B_{n-1,k}\left(1!x_1,2!x_2,\dots,\left(n-k\right)!x_{n-k}\right) \,,
\ee
and
\be\label{eq:Fn}
    F_{n}\left(\rho\right) \coloneqq \frac{C_{n+1}\left(\rho\right)}{C_1\left(\rho\right)} = \frac{r_{\text{h}}^{n}}{n!}f\left(\rho\right)\frac{d^{n}}{d\rho^{n}}\frac{1}{f\left(\rho\right)} \,.
\ee
The first few such polynomials are given below
\be\ba
    {}&\mathcal{P}_1 = 1 \,,\quad \mathcal{P}_2\left(x_1\right) = -x_1 \,,\quad \mathcal{P}_3\left(x_1,x_2\right) = -x_2 +2x_1^2 \,, \\
    &P_4\left(x_1,x_2,x_3\right) = -x_3 +5x_2x_1 -5x_1^3 \,, \\
    &\mathcal{P}_5\left(x_1,x_2,x_3,x_4\right) = -x_4 +6x_3x_1 +3x_2^2 -21x_2x_1^2 +14x_1^4 \,. 
\ea\ee

The expansion of Eq.~\eqref{eq:rhoUvModes} is, in fact, a true expansion of $\varepsilon\left(\upsilon,U\right)$ near $U=0$ for the horizons of interest, with each term in the sum being suppressed compared to its previous term, and converges at least asymptotically. This is due to the fact that we are interested in horizons for which $f\left(\rho\right)$ approaches zero near $\rho=0$ at least as fast as $\rho^{1+\epsilon}$, with $\epsilon>0$. As a result, remembering that $\frac{1}{C_1\left(\rho\right)}=f\left(\rho\right)$, the quantity $\frac{1}{C_1\left(-U\right)^{n}}\mathcal{P}_{n}\left(F_1\left(-U\right),F_2\left(-U\right),\dots,F_{n-1}\left(-U\right)\right)$, goes to zero near $U=0$ at least as fast as $\rho^{n\epsilon}$.

Summarizing, the solution of Eq.~\eqref{eq:rhoUvDef} for $\rho\left(\upsilon,U\right)$, written as an asymptotic expansion near $U=0$, is given by
\be
    \rho\left(\upsilon,U\right) \sim -U+r_{\text{h}}\sum_{n=1}^{\infty}\left(\frac{\upsilon-\upsilon_0}{2r_{\text{h}}}f\left(-U\right)\right)^{n}\mathcal{P}_{n}\left(F_1\left(-U\right),F_2\left(-U\right),\dots,F_{n-1}\left(-U\right)\right) \,,
\ee
where $\mathcal{P}_{n}\left(x_1,x_2,\dots,x_{n}\right)$ is the polynomial constructed as in Eq.~\eqref{eq:PnBnk} and $F_{n}\left(\rho\right)$ defined from derivatives of the metric function $f\left(\rho\right)$ as per Eq.~\eqref{eq:Fn}. Explicitly, the first few terms in the asymptotic expansion are given by
\be\ba\label{eq:rhoUv4}
    {}&\rho\left(\upsilon,U\right) = -U \\
    &+\frac{\upsilon-\upsilon_0}{2}f\left(-U\right)\bigg\{1 +\frac{\upsilon-\upsilon_0}{2}f^{\prime}\left(-U\right) +\frac{1}{2}\left(\frac{\upsilon-\upsilon_0}{2}\right)^2\left[f\left(-U\right)f^{\prime\prime}\left(-U\right) +2f^{\prime2}\left(-U\right)\right] \\
    &+\frac{1}{6}\left(\frac{\upsilon-\upsilon_0}{2}\right)^3\left[f^2\left(-U\right)f^{\prime\prime\prime}\left(-U\right) +9f\left(-U\right)f^{\prime\prime}\left(-U\right)f^{\prime}\left(-U\right) +6f^{\prime3}\left(-U\right)\right] +\dots\bigg\} \,. 
\ea\ee


\section{Numerical Setup}
\label{app:NumericalSetup}

In this appendix we explain the method used for numerical solutions presented in section~\ref{sec:NumericalResults} which follows closely the methods presented in~\cite{Lucietti:2012xr}. After discretizing the $\upsilon$ and $U$ directions into a grid $\left(\upsilon_{a},U_{i}\right)$, with $a=0,\dots,N_{\upsilon}$ and $i=0,\dots,N_{U}$, the evolution equation for the fields, Eq.~\eqref{eq:ODEDoubleNull1}, can be discretized and expressed in terms of $\varphi_{a,i} \coloneqq \varphi(\upsilon_{a},U_{i})$. The differential equation can then be solved for the field using finite difference given by~\footnote{See Appendix A of~\cite{Lucietti:2012xr}.},
\begin{equation}
\begin{split}
    \varphi_{a+1,i+1} &= \left( 1 + \frac{\delta\upsilon_{a}\delta U_{i}}{16} \mathcal V_{a+1, i+1}\right)^{-1}\\& \qquad \left(\varphi_{a+1,i}+\varphi_{a,i+1} - \varphi_{a,i} - \frac{\delta\upsilon_a\delta U_i}{16}\left(\mathcal V_{a+1,i}\varphi_{a+1,i} + \mathcal V_{a,i+1}\varphi_{a,i+1} + \mathcal V_{a,i}\varphi_{a,i} \right) \right) \\
    &\quad+\mathcal{O}\left(\delta\upsilon_{a}\delta U_{i}^3,\delta\upsilon_{a}^3\delta U_{i}\right) \,,
\end{split}
\end{equation}
where $\delta\upsilon_{a}\coloneqq\upsilon_{a+1}-\upsilon_{a}$ and $\delta U_{i}\coloneqq U_{i+1}-U_{i}$.

This is a well-posed initial value problem once $\varphi_{a,0}$ and $\varphi_{0,i}$ are inputted. For the purposes of the current work, we start with either an ingoing or an outgoing wavepacket as initial condition for the field and, together with the effective potential evaluated on the grid, the above equation can be solved numerically to evaluate the field value at the entire $\left(\upsilon_{a},U_{i}\right)$ grid.

For generating the grid in $\upsilon$-direction a uniform scheme was used starting from $\upsilon_0=0$, while the max value $\upsilon_{N_{\upsilon}}$ and grid resolution $N_{\upsilon}$ are reported on the respective parts of Section~\ref{sec:NumericalResults}. In the $U$-direction we used a dynamical grid which gets finer moving closed to horizon located at $U=0$ and spans from $U_0=-0.5$ up to $U_{N_{U}}=0$. We follow a prescription similar to that of~\cite{Lucietti:2012xr} for discretizing the $U$-direction, where the relative distances between the points are given by $\delta U_i = c_i \times \frac{0.5}{\sum_j c_j}$. The distances are computed in terms of $c_i$, given by the scheme $c_{i}=4^{\left\lfloor\frac{5}{N}\left(N-i\right)\right\rfloor}$, where $N$ is the total number of points.

To compute the evolution of the radial derivatives of the field on the horizon, we re-express the equation of motion in $\left(\upsilon,r\right)$ coordinates. For $ds^2=-f\left(\rho\right)\dd\upsilon^2+2\dd\upsilon\dd\rho+r^2\dd\Omega_2^2$, $\Box\Phi=0$ yields for the radiative spherical harmonic modes, $\varphi_{\ell m}\coloneqq \oint_{\mathbb{S}^2}d\Omega_2 \bar{Y}_{\ell m}\,r\Phi$,
\be\ba
    2\partial_{\upsilon}\partial_{\rho}\varphi_{\ell m} &= -f\partial_{\rho}^2\varphi_{\ell m}-f^{\prime}\partial_{\rho}\varphi_{\ell m} +V_{\ell}\left(\rho\right)\varphi_{\ell m} \,,
\ea\ee
where primes denote derivatives w.r.t. $\rho$ and $V_{\ell}\left(\rho\right)$ is the Regge-Wheeler potential for the massless scalar field,
\be
    V_{\ell}\left(\rho\right) = \frac{\ell\left(\ell+1\right)}{r^2}+\frac{\left(fr^{\prime}\right)^{\prime}}{r} \,.
\ee
Acting on the above equation with $\partial_{\rho}$ and setting $\rho = 0$ then successively yields the evolution equation for the field radial derivatives on the horizon. The evolution equation can then be integrated using the trapezoid method. For example, the first few equations are given by,
\begin{equation}
    \begin{split}
        2\partial_{\upsilon}\left(\partial_\rho\varphi_{\ell m}\right)\big|_{\rho=0} &= V_{\ell}\big|_{\rho=0}\ \varphi_{\ell m}\big|_{\rho=0} \,, \\
        2\partial_{\upsilon}\left(\partial_\rho^2 \varphi_{\ell m}\right)\big|_{\rho=0} &= (V_{\ell} - f'')\big|_{\rho=0}\left(\partial_\rho\varphi_{\ell m}\right)\big|_{\rho=0} + V_{\ell}'\big|_{\rho=0} \varphi_{\ell m}\big|_{\rho=0} \\
        2\partial_{\upsilon}\left(\partial_\rho^3 \varphi_{\ell m}\right)\big|_{\rho=0} &= (V_{\ell} - 3 f'')\big|_{\rho=0}\left(\partial_\rho^2\varphi_{\ell m}\right)\big|_{\rho=0} +(2V_{\ell}' - f''')\big|_{\rho=0}\left(\partial_\rho \varphi_{\ell m}\right)\big|_{\rho=0} \\
        &\quad+ V_{\ell}''\big|_{\rho=0}\varphi_{\ell m}\big|_{\rho=0} \,,
    \end{split}
\end{equation}
where we are assuming a degenerate horizon of degeneracy rank at least $2$, i.e. $f|_{\rho=0}=0=f^{\prime}|_{\rho=0}$. More generally, for the evolution of the $k$'th transverse derivative on the horizon,
\be
    2\partial_{\upsilon}\left(\partial_{\rho}^{k}\varphi_{\ell m}\right)\big|_{\rho=0} = \sum_{n=0}^{k-1}F_{\ell;k,n}\left(\partial_{\rho}^{n}\varphi_{\ell m}\right)\big|_{\rho=0} \,,
\ee
with
\be
    F_{\ell;k,n}\coloneqq \left[\binom{k-1}{n}V_{\ell}^{\left(k-n-1\right)} -\binom{k}{n-1}f^{\left(k-n+1\right)}\right]\bigg|_{\rho=0} \,,
\ee
the important observation being that the source terms for the evolution of the $k$'th derivative depend only on the $n\le k-1$ derivatives, thus allowing for an inductive construction of $\left(\partial_{\rho}^{k}\varphi_{\ell m}\right)\big|_{\rho=0}\left(\upsilon\right)$, by inputting only $\varphi_{\ell m}\big|_{\rho=0}\left(\upsilon\right)$.


\addcontentsline{toc}{section}{References}
\bibliographystyle{JHEP}
\bibliography{references}

\end{document}